\documentclass[aps,prl,twocolumn,groupedaddress,floatfix]{revtex4-1}
\usepackage{amsmath}
\usepackage{amssymb}
\usepackage{graphicx}
\usepackage{float}
\usepackage[english]{babel}
\usepackage{dsfont}
\usepackage{epstopdf}
\usepackage{soul}
\usepackage{color}
\usepackage{braket}
\usepackage[normalem]{ulem}

\usepackage[colorlinks=true,
            linkcolor=magenta,
            urlcolor=blue,
            citecolor=blue]{hyperref}

\definecolor{SMblue}{rgb}{0.5,0.8,1}

\emergencystretch=\maxdimen
\usepackage{titlesec}
\titlespacing\subsection{0pt}{12pt plus 4pt minus 2pt}{12pt plus 2pt minus 2pt}

\begin{document}
\title{Observation of unidirectional soliton-like edge states in \\ nonlinear Floquet topological insulators}
\author{Sebabrata~Mukherjee}
\email{mukherjeesebabrata@gmail.com}
\affiliation{Department of Physics, The Pennsylvania State University, University Park, PA 16802, USA}
\author{Mikael C.~Rechtsman}
\email{mcrworld@psu.edu}
\affiliation{Department of Physics, The Pennsylvania State University, University Park, PA 16802, USA}

\date{\today}

\begin{abstract}
\end{abstract}

\maketitle
{\bf
A salient feature of solid-state topological materials in two dimensions is the presence of conducting electronic edge states that are insensitive to scattering by disorder~\cite{klitzing1980new, thouless1982quantized, halperin1982quantized, haldane1988model, kane2005quantum, bernevig2006quantum}.  
Such unidirectional edge states have been explored in many experimental settings beyond solid-state electronic systems, including in photonic devices, mechanical and acoustic structures, and others~\cite{raghu2008analogs, wang2009observation, rechtsman2013photonic, hafezi2013imaging, nash2015topological, susstrunk2015observation, ningyuan2015time, karzig2015topological, nalitov2015polariton, klembt2018exciton, delplace2017topological, ozawa2019topological}. It is of great interest to understand how topological states behave in the presence of inter-particle interactions and nonlinearity~\cite{lumer2013self, ablowitz2014linear, leykam2016edge, ablowitz2017tight, marzuola2019bulk, mukherjee2020observation, smirnova2020nonlinear, xia2020nontrivial, maczewsky2020demonstration, rachel2018interacting, harari2018topological, bandres2018topological}.  
Here we experimentally demonstrate unidirectional soliton-like nonlinear states on the edge of photonic topological insulators consisting of laser-written waveguides. As a result of the optical Kerr nonlinearity of the ambient glass, the soliton-like wavepacket forms a non-diffracting coherent structure that slowly radiates power 
because of the intrinsic gaplessness of the system. 
The realization of soliton-like edge states paves the way to an understanding of topological phenomena in nonlinear systems and those with inter-particle interactions. 
}


The field of topological photonics~\cite{ozawa2019topological} has shown great promise for the discovery of new fundamental science and its implications for advances in optical devices~\cite{hafezi2011robust, bandres2018topological, Guglielmon2019Broadband}.  
Quantum Hall-like topological states for electromagnetic waves were first proposed~\cite{raghu2008analogs} in the context of photonic crystals and were experimentally demonstrated~\cite{wang2009observation} using magneto-optical materials at microwave frequencies.  The concept of `Floquet topological insulators' - namely, inducing topologically nontrivial behavior using dynamical modulation - was used in waveguide arrays for the realization of optical chiral edge states in Chern~\cite{rechtsman2013photonic} and anomalous Floquet topological insulators~\cite{rudner2013anomalous, mukherjee2017experimental, maczewsky2017observation, mukherjee2018state}. In these systems, the component waveguides are spatially modulated along the propagation axis to effectively break time-reversal symmetry in the transverse plane and thus enable the presence of topological gaps and chiral edge states.  Analogous techniques were used to realize topological states in ultracold atomic systems as well \cite{jotzu2014experimental, wintersperger2020realization}. 

\begin{figure*}[t!]
\center
\includegraphics[width=\linewidth]{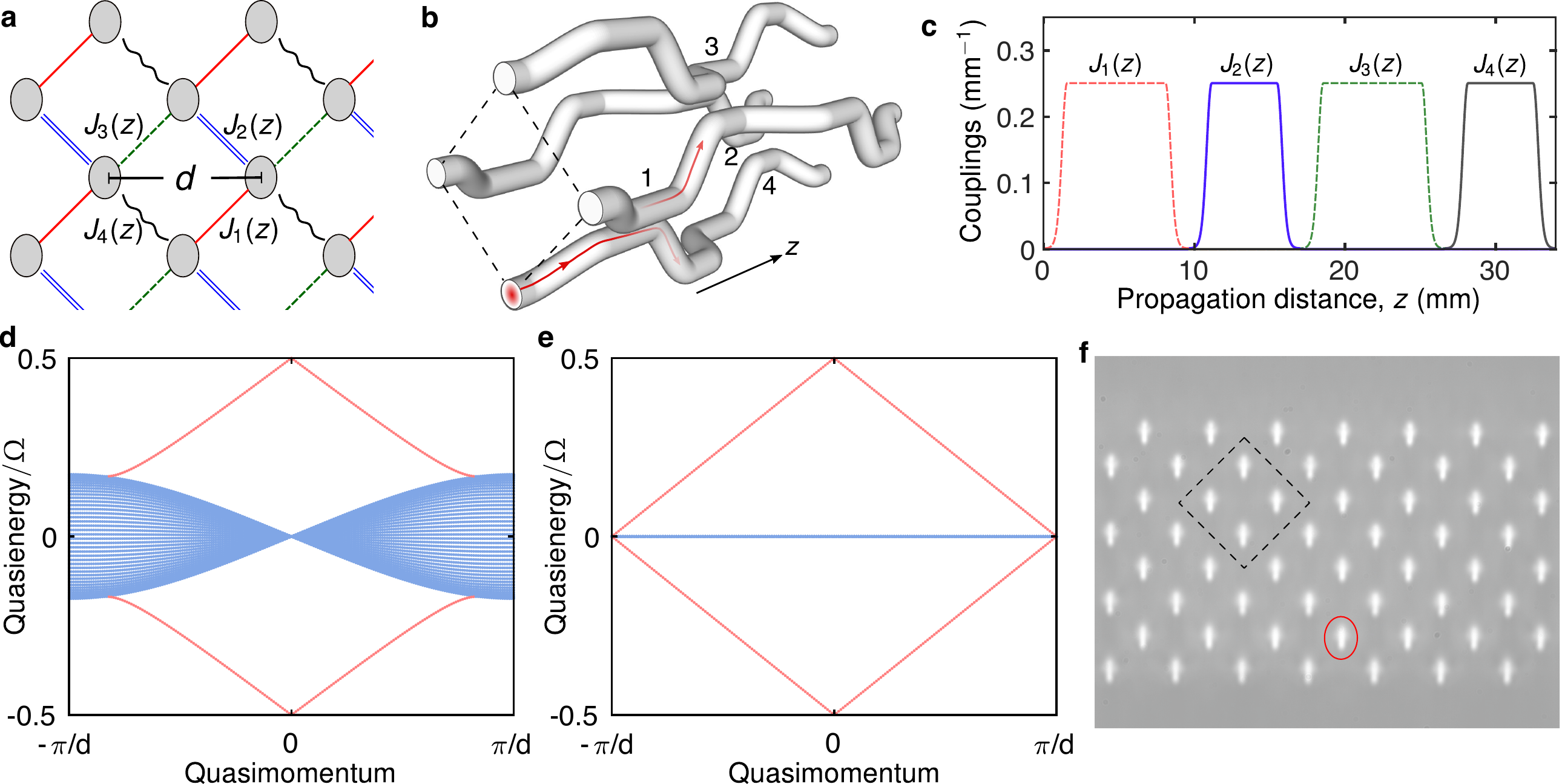}
\caption{{\bf Driving protocol for realizing a photonic Floquet topological insulator and its quasienergy spectrum.} {\bf (a)}  A periodically-driven square lattice where the four couplings $J_m(z)$ [$m\!=\!1,..,4$] are switched on/off in a cyclic (spatially and $z$-periodic) manner. {\bf (b)} Schematic depiction of the implementation of the driving protocol in (a) using a three-dimensional waveguide array. Note that the straight sections of the coupling region have bipartite lengths, depending on $\delta$. {\bf (c)} The couplings resemble a step-like function with a fixed gradual rise and fall `time'. {\bf (d)} Quasienergy spectrum of the bipartite lattice ($\delta\!=\!0.18$) showing two ungapped bulk bands (blue) with zero net Chern number and chiral edge states (red). {\bf (e)} The spectrum in the flat-band limit, i.e., for $\delta\!=\!0$. {\bf (f)} Micrograph of the facet of a periodically-driven square lattice consisting of 56 waveguides. A single-site input state, launched at the circled site on the zig-zag edge, has a significant overlap with the topological edge states. The dashed diamond indicates the four waveguides shown in (b).
}
\label{fig1}
\end{figure*}


Solitons are nonlinear wavepackets that balance nonlinearity with the tendency to spread due to diffraction or dispersion.  
The result is a wavefunction that maintains its shape as it propagates.  Solitons play a central role in the theory of nonlinear differential equations as the solutions that form the basis for the inverse scattering transformation \cite{shabat1972exact, ablowitz1981solitons}, and arise naturally in nonlinear and interacting systems such as water waves \cite{camassa1993integrable}, photonic systems~\cite{barthelemy1985propagation, christodoulides1988discrete, segev1992spatial, eisenberg1998discrete, fleischer2003observation} and Bose-Einstein condensates~\cite{denschlag2000generating, burger1999dark}.  Their ubiquity across different physical platforms speaks to the generality of the equations describing interactions in bosonic systems: the nonlinear diffraction of light and the temporal dynamics of a dense Bose-Einstein condensate are both described by the nonlinear Schr\"odinger equation (also called the Gross-Pitaevskii equation).  The question of how solitons behave in topological systems, both within the bulk and localized to the edge, is an active and open one~\cite{lumer2013self, ablowitz2014linear, leykam2016edge, ablowitz2017tight, marzuola2019bulk, mukherjee2020observation}.

Here, we observe unidirectional soliton-like wavepackets on the edge of a two-dimensional anomalous photonic Floquet topological insulator, where the nonlinearity arises from the optical Kerr effect.  The reason that they must be called `soliton-like' is that they radiate power at a small but non-zero rate.  
Indeed, due to the gaplessness of the system, which arises due to its topological nature, any nonlinear wavepacket would have to radiate energy to the ambient extended modes (with the idiosyncratic exception of embedded solitons \cite{yang1999embedded, yang2001embedded, yagasaki2005discrete}, which is not the case described here).  Furthermore, it is known that moving `solitons' with non-zero velocity in discrete systems have a finite lifetime in general \cite{oxtoby2007moving, kevrekidis2009discrete}.  

We utilize a square array of femtosecond-laser-written optical waveguides, with nearest-neighbor evanescent couplings, where the Kerr nonlinearity increases the refractive index in proportion to the local intensity of light. The lattice is spatially modulated periodically along the propagation direction in a four-step cyclic manner by `switching' the nearest-neighbor couplings on and off -- at a given driving step, every lattice site is coupled to only one of its nearest neighbors, see Figs.~\ref{fig1}~(a, b). In the scalar-paraxial approximation, the propagation of light in the array is governed by the following discrete nonlinear Schr\"odinger equation:
\begin{eqnarray}
\label{nlse}
i\frac{\partial}{\partial z} \Phi_s(z)=\sum_{\left\langle s' \right\rangle} - J_m(z) \Phi_{s'} - |\Phi_s|^2 \Phi_s\; , \label{eq1}
\end{eqnarray}
where the propagation distance $z$ plays the role of time, $s$ labels the lattice sites, and $J_m(z)$ [$m\!=\!1,..,4$] denote the couplings at the $m$-th step. Here, the equation can be rescaled such that $|\Phi_s|^2\!=\! g|\Psi_s|^2$, where $|\Psi_s|^2$ is the optical power at the $s$-th waveguide, and $g\!>\!0$ is determined by the nonlinear refractive index of the medium.
Eq.~\ref{nlse} is mathematically equivalent to the time-dependent Gross-Pitaevskii equation, which is the mean-field description of a Bose-Einstein condensate with attractive interactions. 

\begin{figure*}[t!]
\center
\includegraphics[width=14.5cm]{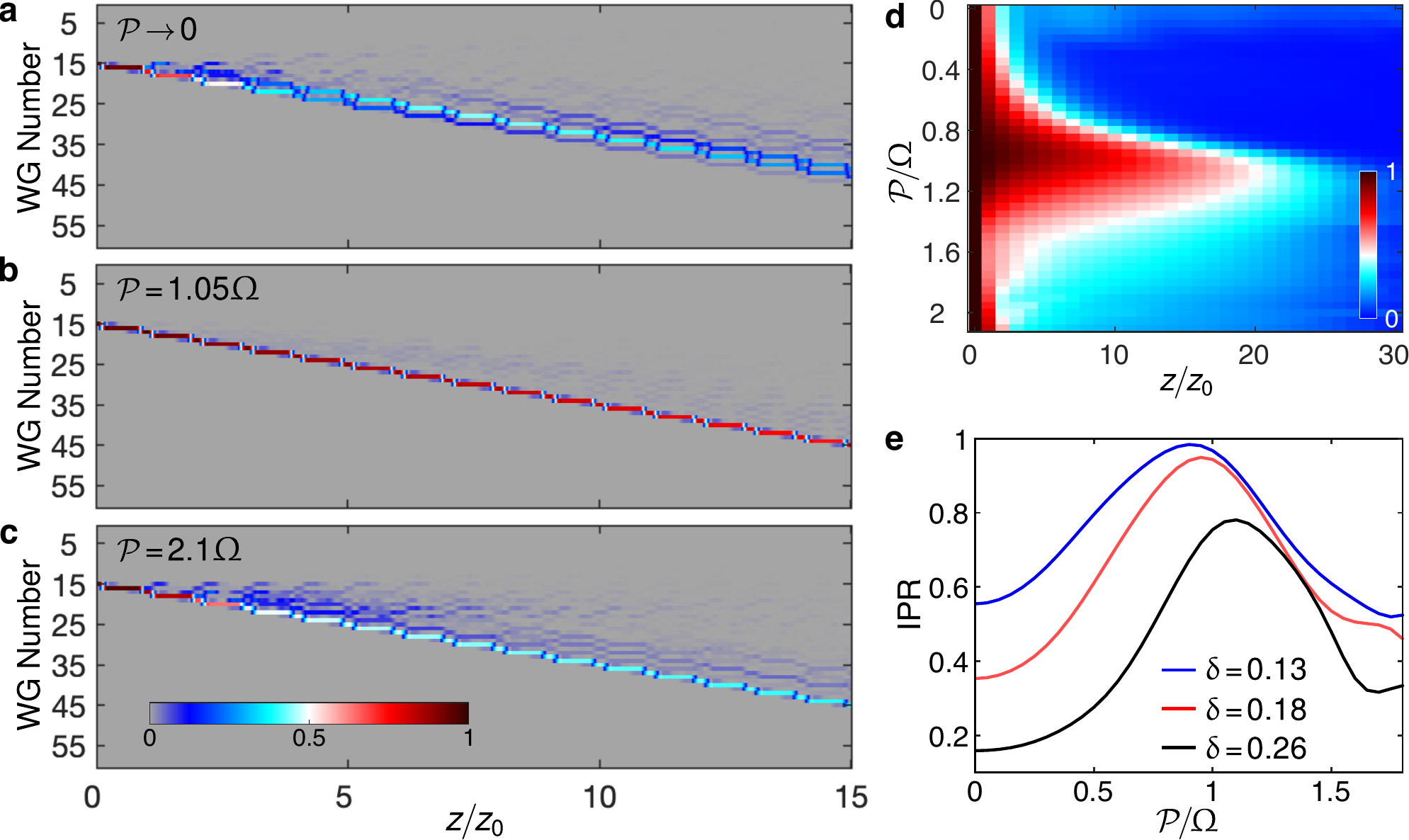}
\caption{{\bf Unidirectional traveling soliton-like edge states. (a-c)} Propagation of  a single-site input wavepacket on the edge of the periodically-driven lattice $(\delta\!=\!0.18)$ with three different renormalized powers, $\mathcal{P}$, indicated on each image. The vertical axis shows the waveguide number along the zig-zag edge. For $\mathcal{P}\!\rightarrow \!0$ (the linear regime) and $\mathcal{P}\!=\!2.1\Omega$ (i.e.~Figs.~a and c), the input state spreads out along the edge as well as penetrates the bulk (not shown here). At a certain intermediate power value (determined by $\delta$), the input state propagates unidirectionally while maintaining its shape (Fig.~b) up to a long propagation distance, see Supplementary Movie 1-3. {\bf(d)} Inverse participation ratio (IPR), calculated after each driving period $z_0\!=\!2\pi/\Omega$ as a function of power and propagation distance. A clear peak in the IPR is observed as a function of $\mathcal{P}$, corresponding to the non-diffracting soliton-like state.  {\bf(e)} Variation of IPR at $z\!=\!2z_0$ for three different values of $\delta$ that were realized experimentally, see Fig.~\ref{fig3}. The IPR at the peak is higher for lower $\delta$, and the value of $\mathcal{P}$ for which the peak occurs increases with $\delta$. 
}
\label{fig2}
\end{figure*}

The complete spatial modulation cycle is divided into four steps. For the $m$-th driving step, we define
$\Lambda_m\!=\! \int {\text{d}} z  J_m(z) $, which determines the transfer of optical power from a given site to one of its nearest neighbors in the linear regime. As shown in Fig.~\ref{fig1}~(c), the couplings $J_m(z)$ resembles a step-like function with a fixed gradual rise and fall ‘time’, however, the driving steps are engineered in a bipartite fashion such that $\Lambda_{1}\!=\!\Lambda_{3}=\pi/2(1+\delta)$ and $\Lambda_{2}\!=\!\Lambda_{4}=\pi/2(1-\delta)$.  The lattice can thus be considered to be effectively `dimerized' with $\delta$ as the degree of dimerization. The linear tight-binding Hamiltonian associated with this model changes periodically in $z$, $H(z+z_0)=H(z)$, with a period $z_0\!=\!2\pi/\Omega$. Note that the driving frequency $\Omega$ remains unaltered when $\delta$ is varied. The quasienergy spectrum of this $z$-periodic lattice system can be obtained by diagonalizing the propagator over one complete period, where the propagator is given by: 
\begin{eqnarray}
\label{EvolutionOperator}
\hat{U}(z_0)\!=\!\mathcal{T} \exp \Big(-i\int_0^{z_0} {\text{d}}\tilde{z} \hat{H}(\tilde{z})  \Big),
\end{eqnarray}
where $\mathcal{T} $ indicates the `time' ordering in $\tilde{z}$. Fig.~\ref{fig1}~(d, e) shows the quasienergy spectrum for $\delta\!=\!0.18$ and $0$, respectively, calculated using a strip geometry aligned along the vertical direction and periodic along the horizontal direction.  The spectrum consists of two ungapped bulk bands (henceforth referred to as the bulk band) and one chiral edge state (per edge) connecting the top and bottom of the bulk band. 

The topological nature of the system, as well as the width of the bulk band, can be controlled by $\delta$. For $\delta\!=\!0$, the bulk band is perfectly dispersionless (flat) and the system is topological; the bulk bandwidth increases with $\delta$, and the bandgap closes at $\delta\!=\!0.5$. Therefore, unidirectional propagating topological edge states exist for $0\! \le \!\delta < \! 0.5$.  
For such a periodically-driven model, the appropriate topological invariant is not the Chern number. Indeed, while chiral edge states are present (see Figs.~\ref{fig1}~(d, e)), the net Chern number of the bulk band is zero -- which would be a violation of the bulk-edge correspondence in an undriven system. 
The topology of this system is captured by another integer-valued invariant known as the Floquet winding number that takes into account the full $z$-evolution~\cite{rudner2013anomalous}, including the micromotion. For a finite lattice with edges, the number of topological edge states present in a gap, which is $1$ for Figs.~\ref{fig1}~(d, e), is directly given by the winding number. 
Such nontrivial lattices, coined {\it anomalous} Floquet topological insulators~\cite{rudner2013anomalous, mukherjee2017experimental, maczewsky2017observation, mukherjee2018state} are unique to periodic driving. Since the quasienergy is periodic, topological edge states can traverse the bandgap around $\Omega/2$, connecting the top and bottom of the bulk bands, see Fig.~\ref{fig1}~(d, e).

\begin{figure*}[t!]
\center
\includegraphics[width=14cm]{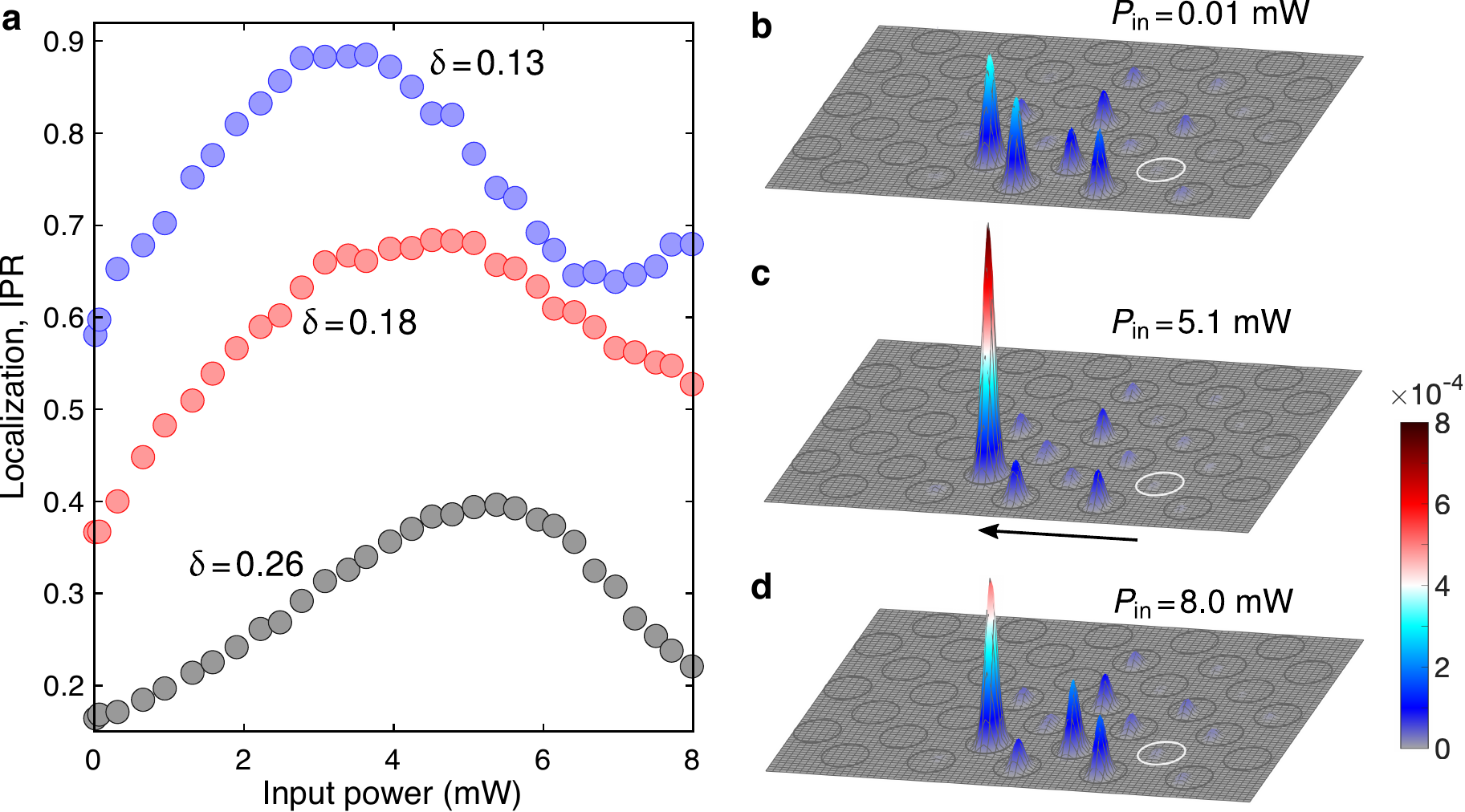}
\caption{{\bf Observation of soliton-like states traveling unidirectionally along the edge of the topological lattice. (a)} Experimentally measured inverse participation ratio at $z=2z_0$ for  $\delta\!=\!\{0.13, 0.18, 0.26\}$ -- for all cases, a clear peak in the IPR is visible. {\bf(b-d)} Output intensity distributions for $\delta\!=\!0.26$ measured at three different power values indicated on each image. The input state propagates two unit cells along the bottom edge with minimal spreading when the average input power is $P_{\text{in}}\!=\!5.1$ mW i.e., (c). The white circle on each image indicates the lattice site where light was launched at the input. 
Each image is normalized, and the field of view is smaller than the lattice size.
}
\label{fig3}
\end{figure*}

In this work, we seek localized, nonlinear traveling edge states that are highly localized on the zig-zag edge (a micrograph of the input facet of the lattice is shown in Fig.~\ref{fig1}~(f)). 
We consider the evolution governed by Eq.~\ref{nlse} with a single-site input state on the edge. The evolution of the normalized optical intensity along the edge is presented in Figs.~\ref{fig2}~(a-c) for $\delta\!=\!0.18$ and three different renormalized powers $(\mathcal{P}\! \equiv \sum_s |\Phi_s|^2)$ indicated on each figure. For $\mathcal{P}\!\rightarrow \!0$ i.e., in the linear regime, the single-site input state has a large overlap $(\approx\!68\%)$ with the topological edge states, see Fig.~\ref{fig2}~(a) and Supplementary Movie~1. Note that the light on the edge diffracts as a result of the curvature of the edge band, see Fig.~\ref{fig1}~(d). This diffraction of the input state can be balanced by nonlinearity, i.e.,~by increasing the optical power. At a certain value of renormalized power (which is an increasing function of $\delta$), the single-site input state propagates unidirectionally without diffraction, one lattice constant per driving period, 
up to a long propagation distance, see Fig.~\ref{fig2}~(b) and Supplementary Movie~2. On the other hand, at higher power, the input state again exhibits a large amount of spreading, as shown in Fig.~\ref{fig2}~(c) and Supplementary Movie~3. 
We note that the behavior shown in Fig.~\ref{fig2}~(b) does not occur when only the edge waveguides are present (i.e., the bulk is removed).

The localization of the state as a function of renormalized power and propagation distance can be quantified by the inverse participation ratio, defined as
\begin{eqnarray}
\label{IPR}
{\text{IPR}}\! = \! \frac{\sum_s |\Phi_s|^4}{\big(\sum_s |\Phi_s|^2\big)^2}\,.
\end{eqnarray}
When all the light is localized at a single site, the inverse participation ratio is at its maximum value of $1$. Fig.~\ref{fig2}~(d) presents the IPR that was calculated stroboscopically, i.e.,~after each driving period for the above-mentioned single-site input state. Up to approximately twenty driving cycles, the IPR exhibits a clear peak as a function of renormalized power at a given propagation distance -- this peak corresponds to soliton-like wavepacket propagation. 
Unlike the intensity distribution along the edge in Figs.~\ref{fig2}~(a-c), the IPR in Fig.~\ref{fig2}~(d) accounts for the entire wavefunction, including the bulk and edge. We note that the localization feature shown in Figs.~\ref{fig2}~(a-d), can be controlled by tuning $\delta$, i.e., by changing the bulk bandwidth. Indeed, the lifetime of the soliton-like edge state shown in Fig.~\ref{fig2}~(b) can be increased arbitrarily by reducing $\delta$ (see Supplementary Materials), and diverges as $\delta\!\rightarrow \!0$, i.e., the flat-band limit. To show how this localization peak depends on $\delta$, we compute the IPR as a function of $\mathcal{P}$ for three different values of $\delta$, namely 0.13, 0.18 and 0.26, associated with experimentally achieved parameters. Fig.~\ref{fig2}~(e) shows the variation of IPR as a function of $\mathcal{P}$ for each value of $\delta$, calculated after a fixed propagation distance $z\!=\!2z_0$.  Here, a clear peak in IPR can be observed for all three values of $\delta$.

Two key observations can be made about the peaks in Fig.~\ref{fig2}~(e): firstly, the IPR at the peak decreases with $\delta$, and secondly, the value of $\mathcal{P}$ at which the peak occurs increases with $\delta$.  These features can be explained as follows.  In the limit of $\delta \! \rightarrow \!0$ (i.e.,~flat-band limit), the edge dispersion is linear and the single-site input state overlaps only with the edge states, hence, this input state 
will propagate along the edge without diffraction in the linear regime ($\mathcal{P}\! \rightarrow \!0$). 
As $\delta$ increases, there is less of an initial overlap on the edge modes, causing the degree of overall localization to decrease as light spreads into the bulk.  Furthermore, the bulk bandwidth increases, implying a faster radiation rate away from the edge and a lower height of the IPR peak.  As power increases (at fixed $\delta$), coupling to bulk modes is induced at first, and then as power is further increased, it acts to trap the wavepacket on the edge, leading to an increase in IPR with power.  At powers past the IPR peak, there is more power than what is necessary to form the localized state, leading to excess diffraction along the edge and into the bulk, and a corresponding decrease in IPR.  The amount of power required to trap the wavepacket along the edge increases with $\delta$, causing a shift in the peak to a higher power with increasing $\delta$.  A detailed explanation of the mechanism of localization can be found in the Supplementary Materials.

We experimentally demonstrate the unidirectional soliton-like edge states by injecting intense laser light into laser-written modulated waveguide arrays with previously mentioned driving protocols.
Here we observe the output states after a fixed propagation distance, as a function of power and for three separate values of $\delta\!=\!\{0.13, 0.18, 0.26\}$. Using femtosecond-laser-writing~\cite{davis1996writing}, three sets of periodically-modulated square lattices, each consisting of $56$ waveguides and with two driving periods, were fabricated inside a $76$-mm-long borosilicate glass substrate. Each waveguide in the lattices supports only the fundamental mode at the operating optical wavelength of $1030$~nm. 
At $z\!=\!0$, the lattice sites are well separated ($27 \, \mu$m inter-site spacing) such that all evanescent couplings are negligibly small. To switch on the coupling between any two neighboring waveguides, the inter-waveguide separation is reduced to $14\, \mu$m, and then the waveguides are kept straight and parallel where the evanescent coupling reaches a fixed and maximal value, see Figs.~\ref{fig1}~(b, c). The couplings are switched off by then separating the waveguides. We control $\delta$ simply by increasing (decreasing) the length of the straight sections of the coupled waveguide pairs in the odd (even) driving step. 

Nonlinear characterization of the photonic lattices was performed using intense laser pulses for which the optical field $\Phi_s$ is a function of both propagation distance and time. The laser pulses may undergo undesired effects such as self-phase modulation (generating new wavelengths) and chromatic dispersion. To access the self-focusing nonlinearity with a minimal self-phase modulation, we use temporally stretched down-chirped laser pulses. 
The effect of chromatic dispersion was estimated to be insignificant for the maximum length scale (i.e., $76$ mm propagation distance) considered here. Additionally, nonlinear loss due to multi-photon absorption was measured to be negligible, see Supplementary Materials.

To experimentally probe the unidirectionally traveling soliton-like edge states, we launch $2$~ps laser pulses into the desired edge waveguide, see Fig.~\ref{fig1}~(e), and then calculate the inverse participation ratio from the measured intensity profile at the output of the lattices. The relationship between the average input power $P_{\text{in}}$ and the renormalized power $\mathcal{P}$ was experimentally determined to be $\mathcal{P} = 0.046$~mm$^{-1}$ per unit $P_{\text{in}}$ in mW; see Supplementary Materials. The nonlinear characterization of the modulated photonic lattices is presented in Fig.~\ref{fig3}~(a), where the measured IPR at $z\!=\!2z_0$ is plotted as a function of the average input power $P_{\text{in}}$ for three different values of $\delta$. 
As a function of $P_{\text{in}}$, the IPR first increases, exhibits a peak at a particular power, and then decreases for higher power -- in all three cases, a clear peak in the IPR is observed, as expected, see Fig.~\ref{fig2}~(e). We show the measured normalized output intensity patterns $|\Phi_s|^2/\mathcal{P}$ for $\delta\!=\!0.26$ in Figs.~\ref{fig3}~(b-d), see also Supplementary Movie~4. Fig.~\ref{fig3}~(b) corresponds to the linear case -- most of the light propagates unidirectionally (i.e., leftward) along the edge, however, a small amount of light penetrates into the bulk. Importantly, note that the light diffracts, i.e., spreads out, along the edge, which is expected from our numerical results. The soliton-like state, corresponding to the IPR peak, is shown in Fig.~\ref{fig3}~(c). Figure~\ref{fig3}~(d) shows the delocalized output intensity at a higher power. 

Comparing Fig.~\ref{fig2}~(e) and Fig.~\ref{fig3}~(a), we observe that the measured IPRs in the linear regime agrees well with the expected values, however, the peaks in the IPRs are experimentally observed at higher powers because of linear losses. Also, the measured heights of the IPR peaks are lower than the expected values, and this effect is more prominent for larger values of $\delta$.  
This lower IPR is caused by both linear losses and a small background due to the linear diffraction of the pulse tails (front and rear) in time.
Having said that, the observed peaks along with the intensity pattern in Fig.~\ref{fig3}~(c) clearly agree with the theoretical predictions of Fig.~\ref{fig2}, and demonstrate the soliton-like edge states in the topological lattice.

In conclusion, we have observed soliton-like edge states propagating unidirectionally along the edge of a nonlinear Floquet topological insulator. This represents a key development in the understanding and use of topological protection against disorder in nonlinear devices.  A natural next step will be to probe the interplay between topological edge states, nonlinearity, and disorder: the fundamental question here is to what extent robustness against disorder carries over to the nonlinear case.  Since the theory of topological materials is linear at its core, we expect that a new theoretical framework will be necessary for describing the nonlinear optics of topological structures.  This will be particularly important if topological protection is to be applied 
to nonlinear applications such as optical switching, on-chip frequency combs and supercontinuum generation, photon entanglement generation and manipulation, and others.      
\\


\noindent{\it Acknowledgments.} The authors acknowledge useful discussions with Daniel Leykam and Jianke Yang. S.M. and M.C.R. gratefully acknowledge support from the Office of Naval Research under award number N00014-18-1-2595, and M.C.R. acknowledges the Packard and Kaufman foundations under fellowship number 2017-66821 and award number KA2017-91788, respectively.

\bibliography{EdgeSol-bib-F}

\begin{thebibliography}{58}%
\makeatletter
\providecommand \@ifxundefined [1]{%
 \@ifx{#1\undefined}
}%
\providecommand \@ifnum [1]{%
 \ifnum #1\expandafter \@firstoftwo
 \else \expandafter \@secondoftwo
 \fi
}%
\providecommand \@ifx [1]{%
 \ifx #1\expandafter \@firstoftwo
 \else \expandafter \@secondoftwo
 \fi
}%
\providecommand \natexlab [1]{#1}%
\providecommand \enquote  [1]{``#1''}%
\providecommand \bibnamefont  [1]{#1}%
\providecommand \bibfnamefont [1]{#1}%
\providecommand \citenamefont [1]{#1}%
\providecommand \href@noop [0]{\@secondoftwo}%
\providecommand \href [0]{\begingroup \@sanitize@url \@href}%
\providecommand \@href[1]{\@@startlink{#1}\@@href}%
\providecommand \@@href[1]{\endgroup#1\@@endlink}%
\providecommand \@sanitize@url [0]{\catcode `\\12\catcode `\$12\catcode
  `\&12\catcode `\#12\catcode `\^12\catcode `\_12\catcode `\%12\relax}%
\providecommand \@@startlink[1]{}%
\providecommand \@@endlink[0]{}%
\providecommand \url  [0]{\begingroup\@sanitize@url \@url }%
\providecommand \@url [1]{\endgroup\@href {#1}{\urlprefix }}%
\providecommand \urlprefix  [0]{URL }%
\providecommand \Eprint [0]{\href }%
\providecommand \doibase [0]{http://dx.doi.org/}%
\providecommand \selectlanguage [0]{\@gobble}%
\providecommand \bibinfo  [0]{\@secondoftwo}%
\providecommand \bibfield  [0]{\@secondoftwo}%
\providecommand \translation [1]{[#1]}%
\providecommand \BibitemOpen [0]{}%
\providecommand \bibitemStop [0]{}%
\providecommand \bibitemNoStop [0]{.\EOS\space}%
\providecommand \EOS [0]{\spacefactor3000\relax}%
\providecommand \BibitemShut  [1]{\csname bibitem#1\endcsname}%
\let\auto@bib@innerbib\@empty
\bibitem [{\citenamefont {Klitzing}\ \emph {et~al.}(1980)\citenamefont
  {Klitzing}, \citenamefont {Dorda},\ and\ \citenamefont
  {Pepper}}]{klitzing1980new}%
  \BibitemOpen
  \bibfield  {author} {\bibinfo {author} {\bibfnamefont {K.~v.}\ \bibnamefont
  {Klitzing}}, \bibinfo {author} {\bibfnamefont {G.}~\bibnamefont {Dorda}}, \
  and\ \bibinfo {author} {\bibfnamefont {M.}~\bibnamefont {Pepper}},\
  }\href@noop {} {\bibfield  {journal} {\bibinfo  {journal} {Phys. Rev. Lett.}\
  }\textbf {\bibinfo {volume} {45}},\ \bibinfo {pages} {494} (\bibinfo {year}
  {1980})}\BibitemShut {NoStop}%
\bibitem [{\citenamefont {Thouless}\ \emph {et~al.}(1982)\citenamefont
  {Thouless}, \citenamefont {Kohmoto}, \citenamefont {Nightingale},\ and\
  \citenamefont {den Nijs}}]{thouless1982quantized}%
  \BibitemOpen
  \bibfield  {author} {\bibinfo {author} {\bibfnamefont {D.~J.}\ \bibnamefont
  {Thouless}}, \bibinfo {author} {\bibfnamefont {M.}~\bibnamefont {Kohmoto}},
  \bibinfo {author} {\bibfnamefont {M.~P.}\ \bibnamefont {Nightingale}}, \ and\
  \bibinfo {author} {\bibfnamefont {M.}~\bibnamefont {den Nijs}},\ }\href@noop
  {} {\bibfield  {journal} {\bibinfo  {journal} {Phys. Rev. Lett.}\ }\textbf
  {\bibinfo {volume} {49}},\ \bibinfo {pages} {405} (\bibinfo {year}
  {1982})}\BibitemShut {NoStop}%
\bibitem [{\citenamefont {Halperin}(1982)}]{halperin1982quantized}%
  \BibitemOpen
  \bibfield  {author} {\bibinfo {author} {\bibfnamefont {B.~I.}\ \bibnamefont
  {Halperin}},\ }\href@noop {} {\bibfield  {journal} {\bibinfo  {journal}
  {Phys. Rev. B}\ }\textbf {\bibinfo {volume} {25}},\ \bibinfo {pages} {2185}
  (\bibinfo {year} {1982})}\BibitemShut {NoStop}%
\bibitem [{\citenamefont {Haldane}(1988)}]{haldane1988model}%
  \BibitemOpen
  \bibfield  {author} {\bibinfo {author} {\bibfnamefont {F.~D.~M.}\
  \bibnamefont {Haldane}},\ }\href@noop {} {\bibfield  {journal} {\bibinfo
  {journal} {Phys. Rev. Lett.}\ }\textbf {\bibinfo {volume} {61}},\ \bibinfo
  {pages} {2015} (\bibinfo {year} {1988})}\BibitemShut {NoStop}%
\bibitem [{\citenamefont {Kane}\ and\ \citenamefont
  {Mele}(2005)}]{kane2005quantum}%
  \BibitemOpen
  \bibfield  {author} {\bibinfo {author} {\bibfnamefont {C.~L.}\ \bibnamefont
  {Kane}}\ and\ \bibinfo {author} {\bibfnamefont {E.~J.}\ \bibnamefont
  {Mele}},\ }\href@noop {} {\bibfield  {journal} {\bibinfo  {journal} {Phys.
  Rev. Lett.}\ }\textbf {\bibinfo {volume} {95}},\ \bibinfo {pages} {226801}
  (\bibinfo {year} {2005})}\BibitemShut {NoStop}%
\bibitem [{\citenamefont {Bernevig}\ \emph {et~al.}(2006)\citenamefont
  {Bernevig}, \citenamefont {Hughes},\ and\ \citenamefont
  {Zhang}}]{bernevig2006quantum}%
  \BibitemOpen
  \bibfield  {author} {\bibinfo {author} {\bibfnamefont {B.~A.}\ \bibnamefont
  {Bernevig}}, \bibinfo {author} {\bibfnamefont {T.~L.}\ \bibnamefont
  {Hughes}}, \ and\ \bibinfo {author} {\bibfnamefont {S.-C.}\ \bibnamefont
  {Zhang}},\ }\href@noop {} {\bibfield  {journal} {\bibinfo  {journal}
  {Science}\ }\textbf {\bibinfo {volume} {314}},\ \bibinfo {pages} {1757}
  (\bibinfo {year} {2006})}\BibitemShut {NoStop}%
\bibitem [{\citenamefont {Raghu}\ and\ \citenamefont
  {Haldane}(2008)}]{raghu2008analogs}%
  \BibitemOpen
  \bibfield  {author} {\bibinfo {author} {\bibfnamefont {S.}~\bibnamefont
  {Raghu}}\ and\ \bibinfo {author} {\bibfnamefont {F.~D.~M.}\ \bibnamefont
  {Haldane}},\ }\href@noop {} {\bibfield  {journal} {\bibinfo  {journal} {Phys.
  Rev. A}\ }\textbf {\bibinfo {volume} {78}},\ \bibinfo {pages} {033834}
  (\bibinfo {year} {2008})}\BibitemShut {NoStop}%
\bibitem [{\citenamefont {Wang}\ \emph {et~al.}(2009)\citenamefont {Wang},
  \citenamefont {Chong}, \citenamefont {Joannopoulos},\ and\ \citenamefont
  {Solja{\v{c}}i{\'c}}}]{wang2009observation}%
  \BibitemOpen
  \bibfield  {author} {\bibinfo {author} {\bibfnamefont {Z.}~\bibnamefont
  {Wang}}, \bibinfo {author} {\bibfnamefont {Y.}~\bibnamefont {Chong}},
  \bibinfo {author} {\bibfnamefont {J.~D.}\ \bibnamefont {Joannopoulos}}, \
  and\ \bibinfo {author} {\bibfnamefont {M.}~\bibnamefont
  {Solja{\v{c}}i{\'c}}},\ }\href@noop {} {\bibfield  {journal} {\bibinfo
  {journal} {Nature}\ }\textbf {\bibinfo {volume} {461}},\ \bibinfo {pages}
  {772} (\bibinfo {year} {2009})}\BibitemShut {NoStop}%
\bibitem [{\citenamefont {Rechtsman}\ \emph {et~al.}(2013)\citenamefont
  {Rechtsman}, \citenamefont {Zeuner}, \citenamefont {Plotnik}, \citenamefont
  {Lumer}, \citenamefont {Podolsky}, \citenamefont {Dreisow}, \citenamefont
  {Nolte}, \citenamefont {Segev},\ and\ \citenamefont
  {Szameit}}]{rechtsman2013photonic}%
  \BibitemOpen
  \bibfield  {author} {\bibinfo {author} {\bibfnamefont {M.~C.}\ \bibnamefont
  {Rechtsman}}, \bibinfo {author} {\bibfnamefont {J.~M.}\ \bibnamefont
  {Zeuner}}, \bibinfo {author} {\bibfnamefont {Y.}~\bibnamefont {Plotnik}},
  \bibinfo {author} {\bibfnamefont {Y.}~\bibnamefont {Lumer}}, \bibinfo
  {author} {\bibfnamefont {D.}~\bibnamefont {Podolsky}}, \bibinfo {author}
  {\bibfnamefont {F.}~\bibnamefont {Dreisow}}, \bibinfo {author} {\bibfnamefont
  {S.}~\bibnamefont {Nolte}}, \bibinfo {author} {\bibfnamefont
  {M.}~\bibnamefont {Segev}}, \ and\ \bibinfo {author} {\bibfnamefont
  {A.}~\bibnamefont {Szameit}},\ }\href@noop {} {\bibfield  {journal} {\bibinfo
   {journal} {Nature}\ }\textbf {\bibinfo {volume} {496}},\ \bibinfo {pages}
  {196} (\bibinfo {year} {2013})}\BibitemShut {NoStop}%
\bibitem [{\citenamefont {Hafezi}\ \emph {et~al.}(2013)\citenamefont {Hafezi},
  \citenamefont {Mittal}, \citenamefont {Fan}, \citenamefont {Migdall},\ and\
  \citenamefont {Taylor}}]{hafezi2013imaging}%
  \BibitemOpen
  \bibfield  {author} {\bibinfo {author} {\bibfnamefont {M.}~\bibnamefont
  {Hafezi}}, \bibinfo {author} {\bibfnamefont {S.}~\bibnamefont {Mittal}},
  \bibinfo {author} {\bibfnamefont {J.}~\bibnamefont {Fan}}, \bibinfo {author}
  {\bibfnamefont {A.}~\bibnamefont {Migdall}}, \ and\ \bibinfo {author}
  {\bibfnamefont {J.}~\bibnamefont {Taylor}},\ }\href@noop {} {\bibfield
  {journal} {\bibinfo  {journal} {Nat. Photonics}\ }\textbf {\bibinfo {volume}
  {7}},\ \bibinfo {pages} {1001} (\bibinfo {year} {2013})}\BibitemShut
  {NoStop}%
\bibitem [{\citenamefont {Nash}\ \emph {et~al.}(2015)\citenamefont {Nash},
  \citenamefont {Kleckner}, \citenamefont {Read}, \citenamefont {Vitelli},
  \citenamefont {Turner},\ and\ \citenamefont {Irvine}}]{nash2015topological}%
  \BibitemOpen
  \bibfield  {author} {\bibinfo {author} {\bibfnamefont {L.~M.}\ \bibnamefont
  {Nash}}, \bibinfo {author} {\bibfnamefont {D.}~\bibnamefont {Kleckner}},
  \bibinfo {author} {\bibfnamefont {A.}~\bibnamefont {Read}}, \bibinfo {author}
  {\bibfnamefont {V.}~\bibnamefont {Vitelli}}, \bibinfo {author} {\bibfnamefont
  {A.~M.}\ \bibnamefont {Turner}}, \ and\ \bibinfo {author} {\bibfnamefont
  {W.~T.}\ \bibnamefont {Irvine}},\ }\href@noop {} {\bibfield  {journal}
  {\bibinfo  {journal} {PNAS}\ }\textbf {\bibinfo {volume} {112}},\ \bibinfo
  {pages} {14495} (\bibinfo {year} {2015})}\BibitemShut {NoStop}%
\bibitem [{\citenamefont {S{\"u}sstrunk}\ and\ \citenamefont
  {Huber}(2015)}]{susstrunk2015observation}%
  \BibitemOpen
  \bibfield  {author} {\bibinfo {author} {\bibfnamefont {R.}~\bibnamefont
  {S{\"u}sstrunk}}\ and\ \bibinfo {author} {\bibfnamefont {S.~D.}\ \bibnamefont
  {Huber}},\ }\href@noop {} {\bibfield  {journal} {\bibinfo  {journal}
  {Science}\ }\textbf {\bibinfo {volume} {349}},\ \bibinfo {pages} {47}
  (\bibinfo {year} {2015})}\BibitemShut {NoStop}%
\bibitem [{\citenamefont {Ningyuan}\ \emph {et~al.}(2015)\citenamefont
  {Ningyuan}, \citenamefont {Owens}, \citenamefont {Sommer}, \citenamefont
  {Schuster},\ and\ \citenamefont {Simon}}]{ningyuan2015time}%
  \BibitemOpen
  \bibfield  {author} {\bibinfo {author} {\bibfnamefont {J.}~\bibnamefont
  {Ningyuan}}, \bibinfo {author} {\bibfnamefont {C.}~\bibnamefont {Owens}},
  \bibinfo {author} {\bibfnamefont {A.}~\bibnamefont {Sommer}}, \bibinfo
  {author} {\bibfnamefont {D.}~\bibnamefont {Schuster}}, \ and\ \bibinfo
  {author} {\bibfnamefont {J.}~\bibnamefont {Simon}},\ }\href@noop {}
  {\bibfield  {journal} {\bibinfo  {journal} {Phys. Rev. X}\ }\textbf {\bibinfo
  {volume} {5}},\ \bibinfo {pages} {021031} (\bibinfo {year}
  {2015})}\BibitemShut {NoStop}%
\bibitem [{\citenamefont {Karzig}\ \emph {et~al.}(2015)\citenamefont {Karzig},
  \citenamefont {Bardyn}, \citenamefont {Lindner},\ and\ \citenamefont
  {Refael}}]{karzig2015topological}%
  \BibitemOpen
  \bibfield  {author} {\bibinfo {author} {\bibfnamefont {T.}~\bibnamefont
  {Karzig}}, \bibinfo {author} {\bibfnamefont {C.-E.}\ \bibnamefont {Bardyn}},
  \bibinfo {author} {\bibfnamefont {N.~H.}\ \bibnamefont {Lindner}}, \ and\
  \bibinfo {author} {\bibfnamefont {G.}~\bibnamefont {Refael}},\ }\href@noop {}
  {\bibfield  {journal} {\bibinfo  {journal} {Phys. Rev. X}\ }\textbf {\bibinfo
  {volume} {5}},\ \bibinfo {pages} {031001} (\bibinfo {year}
  {2015})}\BibitemShut {NoStop}%
\bibitem [{\citenamefont {Nalitov}\ \emph {et~al.}(2015)\citenamefont
  {Nalitov}, \citenamefont {Solnyshkov},\ and\ \citenamefont
  {Malpuech}}]{nalitov2015polariton}%
  \BibitemOpen
  \bibfield  {author} {\bibinfo {author} {\bibfnamefont {A.~V.}\ \bibnamefont
  {Nalitov}}, \bibinfo {author} {\bibfnamefont {D.~D.}\ \bibnamefont
  {Solnyshkov}}, \ and\ \bibinfo {author} {\bibfnamefont {G.}~\bibnamefont
  {Malpuech}},\ }\href@noop {} {\bibfield  {journal} {\bibinfo  {journal}
  {Phys. Rev. Lett.}\ }\textbf {\bibinfo {volume} {114}},\ \bibinfo {pages}
  {116401} (\bibinfo {year} {2015})}\BibitemShut {NoStop}%
\bibitem [{\citenamefont {Klembt}\ \emph {et~al.}(2018)\citenamefont {Klembt},
  \citenamefont {Harder}, \citenamefont {Egorov}, \citenamefont {Winkler},
  \citenamefont {Ge}, \citenamefont {Bandres}, \citenamefont {Emmerling},
  \citenamefont {Worschech}, \citenamefont {Liew}, \citenamefont {Segev} \emph
  {et~al.}}]{klembt2018exciton}%
  \BibitemOpen
  \bibfield  {author} {\bibinfo {author} {\bibfnamefont {S.}~\bibnamefont
  {Klembt}}, \bibinfo {author} {\bibfnamefont {T.~H.}\ \bibnamefont {Harder}},
  \bibinfo {author} {\bibfnamefont {O.~A.}\ \bibnamefont {Egorov}}, \bibinfo
  {author} {\bibfnamefont {K.}~\bibnamefont {Winkler}}, \bibinfo {author}
  {\bibfnamefont {R.}~\bibnamefont {Ge}}, \bibinfo {author} {\bibfnamefont
  {M.~A.}\ \bibnamefont {Bandres}}, \bibinfo {author} {\bibfnamefont
  {M.}~\bibnamefont {Emmerling}}, \bibinfo {author} {\bibfnamefont
  {L.}~\bibnamefont {Worschech}}, \bibinfo {author} {\bibfnamefont {T.~C.~H.}\
  \bibnamefont {Liew}}, \bibinfo {author} {\bibfnamefont {M.}~\bibnamefont
  {Segev}},  \emph {et~al.},\ }\href@noop {} {\bibfield  {journal} {\bibinfo
  {journal} {Nature}\ }\textbf {\bibinfo {volume} {562}},\ \bibinfo {pages}
  {552} (\bibinfo {year} {2018})}\BibitemShut {NoStop}%
\bibitem [{\citenamefont {Delplace}\ \emph {et~al.}(2017)\citenamefont
  {Delplace}, \citenamefont {Marston},\ and\ \citenamefont
  {Venaille}}]{delplace2017topological}%
  \BibitemOpen
  \bibfield  {author} {\bibinfo {author} {\bibfnamefont {P.}~\bibnamefont
  {Delplace}}, \bibinfo {author} {\bibfnamefont {J.~B.}\ \bibnamefont
  {Marston}}, \ and\ \bibinfo {author} {\bibfnamefont {A.}~\bibnamefont
  {Venaille}},\ }\href@noop {} {\bibfield  {journal} {\bibinfo  {journal}
  {Science}\ }\textbf {\bibinfo {volume} {358}},\ \bibinfo {pages} {1075}
  (\bibinfo {year} {2017})}\BibitemShut {NoStop}%
\bibitem [{\citenamefont {Ozawa}\ \emph {et~al.}(2019)\citenamefont {Ozawa},
  \citenamefont {Price}, \citenamefont {Amo}, \citenamefont {Goldman},
  \citenamefont {Hafezi}, \citenamefont {Lu}, \citenamefont {Rechtsman},
  \citenamefont {Schuster}, \citenamefont {Simon}, \citenamefont {Zilberberg}
  \emph {et~al.}}]{ozawa2019topological}%
  \BibitemOpen
  \bibfield  {author} {\bibinfo {author} {\bibfnamefont {T.}~\bibnamefont
  {Ozawa}}, \bibinfo {author} {\bibfnamefont {H.~M.}\ \bibnamefont {Price}},
  \bibinfo {author} {\bibfnamefont {A.}~\bibnamefont {Amo}}, \bibinfo {author}
  {\bibfnamefont {N.}~\bibnamefont {Goldman}}, \bibinfo {author} {\bibfnamefont
  {M.}~\bibnamefont {Hafezi}}, \bibinfo {author} {\bibfnamefont
  {L.}~\bibnamefont {Lu}}, \bibinfo {author} {\bibfnamefont {M.~C.}\
  \bibnamefont {Rechtsman}}, \bibinfo {author} {\bibfnamefont {D.}~\bibnamefont
  {Schuster}}, \bibinfo {author} {\bibfnamefont {J.}~\bibnamefont {Simon}},
  \bibinfo {author} {\bibfnamefont {O.}~\bibnamefont {Zilberberg}},  \emph
  {et~al.},\ }\href@noop {} {\bibfield  {journal} {\bibinfo  {journal} {Rev.
  Mod. Phys.}\ }\textbf {\bibinfo {volume} {91}},\ \bibinfo {pages} {015006}
  (\bibinfo {year} {2019})}\BibitemShut {NoStop}%
\bibitem [{\citenamefont {Lumer}\ \emph {et~al.}(2013)\citenamefont {Lumer},
  \citenamefont {Plotnik}, \citenamefont {Rechtsman},\ and\ \citenamefont
  {Segev}}]{lumer2013self}%
  \BibitemOpen
  \bibfield  {author} {\bibinfo {author} {\bibfnamefont {Y.}~\bibnamefont
  {Lumer}}, \bibinfo {author} {\bibfnamefont {Y.}~\bibnamefont {Plotnik}},
  \bibinfo {author} {\bibfnamefont {M.~C.}\ \bibnamefont {Rechtsman}}, \ and\
  \bibinfo {author} {\bibfnamefont {M.}~\bibnamefont {Segev}},\ }\href@noop {}
  {\bibfield  {journal} {\bibinfo  {journal} {Phys. Rev. Lett.}\ }\textbf
  {\bibinfo {volume} {111}},\ \bibinfo {pages} {243905} (\bibinfo {year}
  {2013})}\BibitemShut {NoStop}%
\bibitem [{\citenamefont {Ablowitz}\ \emph {et~al.}(2014)\citenamefont
  {Ablowitz}, \citenamefont {Curtis},\ and\ \citenamefont
  {Ma}}]{ablowitz2014linear}%
  \BibitemOpen
  \bibfield  {author} {\bibinfo {author} {\bibfnamefont {M.~J.}\ \bibnamefont
  {Ablowitz}}, \bibinfo {author} {\bibfnamefont {C.~W.}\ \bibnamefont
  {Curtis}}, \ and\ \bibinfo {author} {\bibfnamefont {Y.-P.}\ \bibnamefont
  {Ma}},\ }\href@noop {} {\bibfield  {journal} {\bibinfo  {journal} {Phys. Rev.
  A}\ }\textbf {\bibinfo {volume} {90}},\ \bibinfo {pages} {023813} (\bibinfo
  {year} {2014})}\BibitemShut {NoStop}%
\bibitem [{\citenamefont {Leykam}\ and\ \citenamefont
  {Chong}(2016)}]{leykam2016edge}%
  \BibitemOpen
  \bibfield  {author} {\bibinfo {author} {\bibfnamefont {D.}~\bibnamefont
  {Leykam}}\ and\ \bibinfo {author} {\bibfnamefont {Y.~D.}\ \bibnamefont
  {Chong}},\ }\href@noop {} {\bibfield  {journal} {\bibinfo  {journal} {Phys.
  Rev. Lett.}\ }\textbf {\bibinfo {volume} {117}},\ \bibinfo {pages} {143901}
  (\bibinfo {year} {2016})}\BibitemShut {NoStop}%
\bibitem [{\citenamefont {Ablowitz}\ and\ \citenamefont
  {Cole}(2017)}]{ablowitz2017tight}%
  \BibitemOpen
  \bibfield  {author} {\bibinfo {author} {\bibfnamefont {M.~J.}\ \bibnamefont
  {Ablowitz}}\ and\ \bibinfo {author} {\bibfnamefont {J.~T.}\ \bibnamefont
  {Cole}},\ }\href@noop {} {\bibfield  {journal} {\bibinfo  {journal} {Phys.
  Rev. A}\ }\textbf {\bibinfo {volume} {96}},\ \bibinfo {pages} {043868}
  (\bibinfo {year} {2017})}\BibitemShut {NoStop}%
\bibitem [{\citenamefont {Marzuola}\ \emph {et~al.}(2019)\citenamefont
  {Marzuola}, \citenamefont {Rechtsman}, \citenamefont {Osting},\ and\
  \citenamefont {Bandres}}]{marzuola2019bulk}%
  \BibitemOpen
  \bibfield  {author} {\bibinfo {author} {\bibfnamefont {J.~L.}\ \bibnamefont
  {Marzuola}}, \bibinfo {author} {\bibfnamefont {M.}~\bibnamefont {Rechtsman}},
  \bibinfo {author} {\bibfnamefont {B.}~\bibnamefont {Osting}}, \ and\ \bibinfo
  {author} {\bibfnamefont {M.}~\bibnamefont {Bandres}},\ }\href@noop {}
  {\bibfield  {journal} {\bibinfo  {journal} {arXiv preprint arXiv:1904.10312}\
  } (\bibinfo {year} {2019})}\BibitemShut {NoStop}%
\bibitem [{\citenamefont {Mukherjee}\ and\ \citenamefont
  {Rechtsman}(2020)}]{mukherjee2020observation}%
  \BibitemOpen
  \bibfield  {author} {\bibinfo {author} {\bibfnamefont {S.}~\bibnamefont
  {Mukherjee}}\ and\ \bibinfo {author} {\bibfnamefont {M.~C.}\ \bibnamefont
  {Rechtsman}},\ }\href@noop {} {\bibfield  {journal} {\bibinfo  {journal}
  {Science}\ }\textbf {\bibinfo {volume} {368}},\ \bibinfo {pages} {856}
  (\bibinfo {year} {2020})}\BibitemShut {NoStop}%
\bibitem [{\citenamefont {Smirnova}\ \emph {et~al.}(2020)\citenamefont
  {Smirnova}, \citenamefont {Leykam}, \citenamefont {Chong},\ and\
  \citenamefont {Kivshar}}]{smirnova2020nonlinear}%
  \BibitemOpen
  \bibfield  {author} {\bibinfo {author} {\bibfnamefont {D.}~\bibnamefont
  {Smirnova}}, \bibinfo {author} {\bibfnamefont {D.}~\bibnamefont {Leykam}},
  \bibinfo {author} {\bibfnamefont {Y.}~\bibnamefont {Chong}}, \ and\ \bibinfo
  {author} {\bibfnamefont {Y.}~\bibnamefont {Kivshar}},\ }\href@noop {}
  {\bibfield  {journal} {\bibinfo  {journal} {Appl. Phys. Rev.}\ }\textbf
  {\bibinfo {volume} {7}},\ \bibinfo {pages} {021306} (\bibinfo {year}
  {2020})}\BibitemShut {NoStop}%
\bibitem [{\citenamefont {Xia}\ \emph {et~al.}(2020)\citenamefont {Xia},
  \citenamefont {Juki{\'c}}, \citenamefont {Wang}, \citenamefont {Smirnova},
  \citenamefont {Smirnov}, \citenamefont {Tang}, \citenamefont {Song},
  \citenamefont {Szameit}, \citenamefont {Leykam}, \citenamefont {Xu},
  \citenamefont {Chen},\ and\ \citenamefont {Buljan}}]{xia2020nontrivial}%
  \BibitemOpen
  \bibfield  {author} {\bibinfo {author} {\bibfnamefont {S.}~\bibnamefont
  {Xia}}, \bibinfo {author} {\bibfnamefont {D.}~\bibnamefont {Juki{\'c}}},
  \bibinfo {author} {\bibfnamefont {N.}~\bibnamefont {Wang}}, \bibinfo {author}
  {\bibfnamefont {D.}~\bibnamefont {Smirnova}}, \bibinfo {author}
  {\bibfnamefont {L.}~\bibnamefont {Smirnov}}, \bibinfo {author} {\bibfnamefont
  {L.}~\bibnamefont {Tang}}, \bibinfo {author} {\bibfnamefont {D.}~\bibnamefont
  {Song}}, \bibinfo {author} {\bibfnamefont {A.}~\bibnamefont {Szameit}},
  \bibinfo {author} {\bibfnamefont {D.}~\bibnamefont {Leykam}}, \bibinfo
  {author} {\bibfnamefont {J.}~\bibnamefont {Xu}}, \bibinfo {author}
  {\bibfnamefont {Z.}~\bibnamefont {Chen}}, \ and\ \bibinfo {author}
  {\bibfnamefont {H.}~\bibnamefont {Buljan}},\ }\href@noop {} {\bibfield
  {journal} {\bibinfo  {journal} {Light Sci. Appl.}\ }\textbf {\bibinfo
  {volume} {9}},\ \bibinfo {pages} {147} (\bibinfo {year} {2020})}\BibitemShut
  {NoStop}%
\bibitem [{\citenamefont {Maczewsky}\ \emph {et~al.}(2020)\citenamefont
  {Maczewsky}, \citenamefont {Heinrich}, \citenamefont {Kremer}, \citenamefont
  {Ivanov}, \citenamefont {Ehrhardt}, \citenamefont {Martinez}, \citenamefont
  {Kartashov}, \citenamefont {Konotop}, \citenamefont {Torner},\ and\
  \citenamefont {Szameit}}]{maczewsky2020demonstration}%
  \BibitemOpen
  \bibfield  {author} {\bibinfo {author} {\bibfnamefont {L.~J.}\ \bibnamefont
  {Maczewsky}}, \bibinfo {author} {\bibfnamefont {M.}~\bibnamefont {Heinrich}},
  \bibinfo {author} {\bibfnamefont {M.}~\bibnamefont {Kremer}}, \bibinfo
  {author} {\bibfnamefont {S.~K.}\ \bibnamefont {Ivanov}}, \bibinfo {author}
  {\bibfnamefont {M.}~\bibnamefont {Ehrhardt}}, \bibinfo {author}
  {\bibfnamefont {F.}~\bibnamefont {Martinez}}, \bibinfo {author}
  {\bibfnamefont {Y.~V.}\ \bibnamefont {Kartashov}}, \bibinfo {author}
  {\bibfnamefont {V.~V.}\ \bibnamefont {Konotop}}, \bibinfo {author}
  {\bibfnamefont {L.}~\bibnamefont {Torner}}, \ and\ \bibinfo {author}
  {\bibfnamefont {A.}~\bibnamefont {Szameit}},\ }in\ \href@noop {} {\emph
  {\bibinfo {booktitle} {CLEO: QELS\_Fundamental Science}}}\ (\bibinfo
  {organization} {Optical Society of America},\ \bibinfo {year} {2020})\ pp.\
  \bibinfo {pages} {FW4A--2}\BibitemShut {NoStop}%
\bibitem [{\citenamefont {Rachel}(2018)}]{rachel2018interacting}%
  \BibitemOpen
  \bibfield  {author} {\bibinfo {author} {\bibfnamefont {S.}~\bibnamefont
  {Rachel}},\ }\href@noop {} {\bibfield  {journal} {\bibinfo  {journal} {Rep.
  Prog. Phys.}\ }\textbf {\bibinfo {volume} {81}},\ \bibinfo {pages} {116501}
  (\bibinfo {year} {2018})}\BibitemShut {NoStop}%
\bibitem [{\citenamefont {Harari}\ \emph {et~al.}(2018)\citenamefont {Harari},
  \citenamefont {Bandres}, \citenamefont {Lumer}, \citenamefont {Rechtsman},
  \citenamefont {Chong}, \citenamefont {Khajavikhan}, \citenamefont
  {Christodoulides},\ and\ \citenamefont {Segev}}]{harari2018topological}%
  \BibitemOpen
  \bibfield  {author} {\bibinfo {author} {\bibfnamefont {G.}~\bibnamefont
  {Harari}}, \bibinfo {author} {\bibfnamefont {M.~A.}\ \bibnamefont {Bandres}},
  \bibinfo {author} {\bibfnamefont {Y.}~\bibnamefont {Lumer}}, \bibinfo
  {author} {\bibfnamefont {M.~C.}\ \bibnamefont {Rechtsman}}, \bibinfo {author}
  {\bibfnamefont {Y.~D.}\ \bibnamefont {Chong}}, \bibinfo {author}
  {\bibfnamefont {M.}~\bibnamefont {Khajavikhan}}, \bibinfo {author}
  {\bibfnamefont {D.~N.}\ \bibnamefont {Christodoulides}}, \ and\ \bibinfo
  {author} {\bibfnamefont {M.}~\bibnamefont {Segev}},\ }\href@noop {}
  {\bibfield  {journal} {\bibinfo  {journal} {Science}\ }\textbf {\bibinfo
  {volume} {359}},\ \bibinfo {pages} {eaar4003} (\bibinfo {year}
  {2018})}\BibitemShut {NoStop}%
\bibitem [{\citenamefont {Bandres}\ \emph {et~al.}(2018)\citenamefont
  {Bandres}, \citenamefont {Wittek}, \citenamefont {Harari}, \citenamefont
  {Parto}, \citenamefont {Ren}, \citenamefont {Segev}, \citenamefont
  {Christodoulides},\ and\ \citenamefont
  {Khajavikhan}}]{bandres2018topological}%
  \BibitemOpen
  \bibfield  {author} {\bibinfo {author} {\bibfnamefont {M.~A.}\ \bibnamefont
  {Bandres}}, \bibinfo {author} {\bibfnamefont {S.}~\bibnamefont {Wittek}},
  \bibinfo {author} {\bibfnamefont {G.}~\bibnamefont {Harari}}, \bibinfo
  {author} {\bibfnamefont {M.}~\bibnamefont {Parto}}, \bibinfo {author}
  {\bibfnamefont {J.}~\bibnamefont {Ren}}, \bibinfo {author} {\bibfnamefont
  {M.}~\bibnamefont {Segev}}, \bibinfo {author} {\bibfnamefont {D.~N.}\
  \bibnamefont {Christodoulides}}, \ and\ \bibinfo {author} {\bibfnamefont
  {M.}~\bibnamefont {Khajavikhan}},\ }\href@noop {} {\bibfield  {journal}
  {\bibinfo  {journal} {Science}\ }\textbf {\bibinfo {volume} {359}},\ \bibinfo
  {pages} {eaar4005} (\bibinfo {year} {2018})}\BibitemShut {NoStop}%
\bibitem [{\citenamefont {Hafezi}\ \emph {et~al.}(2011)\citenamefont {Hafezi},
  \citenamefont {Demler}, \citenamefont {Lukin},\ and\ \citenamefont
  {Taylor}}]{hafezi2011robust}%
  \BibitemOpen
  \bibfield  {author} {\bibinfo {author} {\bibfnamefont {M.}~\bibnamefont
  {Hafezi}}, \bibinfo {author} {\bibfnamefont {E.~A.}\ \bibnamefont {Demler}},
  \bibinfo {author} {\bibfnamefont {M.~D.}\ \bibnamefont {Lukin}}, \ and\
  \bibinfo {author} {\bibfnamefont {J.~M.}\ \bibnamefont {Taylor}},\
  }\href@noop {} {\bibfield  {journal} {\bibinfo  {journal} {Nat. Phys.}\
  }\textbf {\bibinfo {volume} {7}},\ \bibinfo {pages} {907} (\bibinfo {year}
  {2011})}\BibitemShut {NoStop}%
\bibitem [{\citenamefont {Guglielmon}\ and\ \citenamefont
  {Rechtsman}(2019)}]{Guglielmon2019Broadband}%
  \BibitemOpen
  \bibfield  {author} {\bibinfo {author} {\bibfnamefont {J.}~\bibnamefont
  {Guglielmon}}\ and\ \bibinfo {author} {\bibfnamefont {M.~C.}\ \bibnamefont
  {Rechtsman}},\ }\href@noop {} {\bibfield  {journal} {\bibinfo  {journal}
  {Phys. Rev. Lett.}\ }\textbf {\bibinfo {volume} {122}},\ \bibinfo {pages}
  {153904} (\bibinfo {year} {2019})}\BibitemShut {NoStop}%
\bibitem [{\citenamefont {Rudner}\ \emph {et~al.}(2013)\citenamefont {Rudner},
  \citenamefont {Lindner}, \citenamefont {Berg},\ and\ \citenamefont
  {Levin}}]{rudner2013anomalous}%
  \BibitemOpen
  \bibfield  {author} {\bibinfo {author} {\bibfnamefont {M.~S.}\ \bibnamefont
  {Rudner}}, \bibinfo {author} {\bibfnamefont {N.~H.}\ \bibnamefont {Lindner}},
  \bibinfo {author} {\bibfnamefont {E.}~\bibnamefont {Berg}}, \ and\ \bibinfo
  {author} {\bibfnamefont {M.}~\bibnamefont {Levin}},\ }\href@noop {}
  {\bibfield  {journal} {\bibinfo  {journal} {Phys. Rev. X}\ }\textbf {\bibinfo
  {volume} {3}},\ \bibinfo {pages} {031005} (\bibinfo {year}
  {2013})}\BibitemShut {NoStop}%
\bibitem [{\citenamefont {Mukherjee}\ \emph {et~al.}(2017)\citenamefont
  {Mukherjee}, \citenamefont {Spracklen}, \citenamefont {Valiente},
  \citenamefont {Andersson}, \citenamefont {{\"O}hberg}, \citenamefont
  {Goldman},\ and\ \citenamefont {Thomson}}]{mukherjee2017experimental}%
  \BibitemOpen
  \bibfield  {author} {\bibinfo {author} {\bibfnamefont {S.}~\bibnamefont
  {Mukherjee}}, \bibinfo {author} {\bibfnamefont {A.}~\bibnamefont
  {Spracklen}}, \bibinfo {author} {\bibfnamefont {M.}~\bibnamefont {Valiente}},
  \bibinfo {author} {\bibfnamefont {E.}~\bibnamefont {Andersson}}, \bibinfo
  {author} {\bibfnamefont {P.}~\bibnamefont {{\"O}hberg}}, \bibinfo {author}
  {\bibfnamefont {N.}~\bibnamefont {Goldman}}, \ and\ \bibinfo {author}
  {\bibfnamefont {R.~R.}\ \bibnamefont {Thomson}},\ }\href@noop {} {\bibfield
  {journal} {\bibinfo  {journal} {Nat. Commun.}\ }\textbf {\bibinfo {volume}
  {8}},\ \bibinfo {pages} {13918} (\bibinfo {year} {2017})}\BibitemShut
  {NoStop}%
\bibitem [{\citenamefont {Maczewsky}\ \emph {et~al.}(2017)\citenamefont
  {Maczewsky}, \citenamefont {Zeuner}, \citenamefont {Nolte},\ and\
  \citenamefont {Szameit}}]{maczewsky2017observation}%
  \BibitemOpen
  \bibfield  {author} {\bibinfo {author} {\bibfnamefont {L.~J.}\ \bibnamefont
  {Maczewsky}}, \bibinfo {author} {\bibfnamefont {J.~M.}\ \bibnamefont
  {Zeuner}}, \bibinfo {author} {\bibfnamefont {S.}~\bibnamefont {Nolte}}, \
  and\ \bibinfo {author} {\bibfnamefont {A.}~\bibnamefont {Szameit}},\
  }\href@noop {} {\bibfield  {journal} {\bibinfo  {journal} {Nat. Commun.}\
  }\textbf {\bibinfo {volume} {8}},\ \bibinfo {pages} {13756} (\bibinfo {year}
  {2017})}\BibitemShut {NoStop}%
\bibitem [{\citenamefont {Mukherjee}\ \emph {et~al.}(2018)\citenamefont
  {Mukherjee}, \citenamefont {Chandrasekharan}, \citenamefont {{\"O}hberg},
  \citenamefont {Goldman},\ and\ \citenamefont {Thomson}}]{mukherjee2018state}%
  \BibitemOpen
  \bibfield  {author} {\bibinfo {author} {\bibfnamefont {S.}~\bibnamefont
  {Mukherjee}}, \bibinfo {author} {\bibfnamefont {H.~K.}\ \bibnamefont
  {Chandrasekharan}}, \bibinfo {author} {\bibfnamefont {P.}~\bibnamefont
  {{\"O}hberg}}, \bibinfo {author} {\bibfnamefont {N.}~\bibnamefont {Goldman}},
  \ and\ \bibinfo {author} {\bibfnamefont {R.~R.}\ \bibnamefont {Thomson}},\
  }\href@noop {} {\bibfield  {journal} {\bibinfo  {journal} {Nat. Commun.}\
  }\textbf {\bibinfo {volume} {9}},\ \bibinfo {pages} {4209} (\bibinfo {year}
  {2018})}\BibitemShut {NoStop}%
\bibitem [{\citenamefont {Jotzu}\ \emph {et~al.}(2014)\citenamefont {Jotzu},
  \citenamefont {Messer}, \citenamefont {Desbuquois}, \citenamefont {Lebrat},
  \citenamefont {Uehlinger}, \citenamefont {Greif},\ and\ \citenamefont
  {Esslinger}}]{jotzu2014experimental}%
  \BibitemOpen
  \bibfield  {author} {\bibinfo {author} {\bibfnamefont {G.}~\bibnamefont
  {Jotzu}}, \bibinfo {author} {\bibfnamefont {M.}~\bibnamefont {Messer}},
  \bibinfo {author} {\bibfnamefont {R.}~\bibnamefont {Desbuquois}}, \bibinfo
  {author} {\bibfnamefont {M.}~\bibnamefont {Lebrat}}, \bibinfo {author}
  {\bibfnamefont {T.}~\bibnamefont {Uehlinger}}, \bibinfo {author}
  {\bibfnamefont {D.}~\bibnamefont {Greif}}, \ and\ \bibinfo {author}
  {\bibfnamefont {T.}~\bibnamefont {Esslinger}},\ }\href@noop {} {\bibfield
  {journal} {\bibinfo  {journal} {Nature}\ }\textbf {\bibinfo {volume} {515}},\
  \bibinfo {pages} {237} (\bibinfo {year} {2014})}\BibitemShut {NoStop}%
\bibitem [{\citenamefont {Wintersperger}\ \emph {et~al.}(2020)\citenamefont
  {Wintersperger}, \citenamefont {Braun}, \citenamefont {{\"U}nal},
  \citenamefont {Eckardt}, \citenamefont {Di~Liberto}, \citenamefont {Goldman},
  \citenamefont {Bloch},\ and\ \citenamefont
  {Aidelsburger}}]{wintersperger2020realization}%
  \BibitemOpen
  \bibfield  {author} {\bibinfo {author} {\bibfnamefont {K.}~\bibnamefont
  {Wintersperger}}, \bibinfo {author} {\bibfnamefont {C.}~\bibnamefont
  {Braun}}, \bibinfo {author} {\bibfnamefont {F.~N.}\ \bibnamefont {{\"U}nal}},
  \bibinfo {author} {\bibfnamefont {A.}~\bibnamefont {Eckardt}}, \bibinfo
  {author} {\bibfnamefont {M.}~\bibnamefont {Di~Liberto}}, \bibinfo {author}
  {\bibfnamefont {N.}~\bibnamefont {Goldman}}, \bibinfo {author} {\bibfnamefont
  {I.}~\bibnamefont {Bloch}}, \ and\ \bibinfo {author} {\bibfnamefont
  {M.}~\bibnamefont {Aidelsburger}},\ }\href@noop {} {\bibfield  {journal}
  {\bibinfo  {journal} {Nat. Phys.}\ }\textbf {\bibinfo {volume} {16}},\
  \bibinfo {pages} {1058} (\bibinfo {year} {2020})}\BibitemShut {NoStop}%
\bibitem [{\citenamefont {Shabat}\ and\ \citenamefont
  {Zakharov}(1972)}]{shabat1972exact}%
  \BibitemOpen
  \bibfield  {author} {\bibinfo {author} {\bibfnamefont {A.}~\bibnamefont
  {Shabat}}\ and\ \bibinfo {author} {\bibfnamefont {V.}~\bibnamefont
  {Zakharov}},\ }\href@noop {} {\bibfield  {journal} {\bibinfo  {journal}
  {Soviet physics JETP}\ }\textbf {\bibinfo {volume} {34}},\ \bibinfo {pages}
  {62} (\bibinfo {year} {1972})}\BibitemShut {NoStop}%
\bibitem [{\citenamefont {Ablowitz}\ and\ \citenamefont
  {Segur}(1981)}]{ablowitz1981solitons}%
  \BibitemOpen
  \bibfield  {author} {\bibinfo {author} {\bibfnamefont {M.~J.}\ \bibnamefont
  {Ablowitz}}\ and\ \bibinfo {author} {\bibfnamefont {H.}~\bibnamefont
  {Segur}},\ }\href@noop {} {\emph {\bibinfo {title} {Solitons and the inverse
  scattering transform}}}\ (\bibinfo  {publisher} {SIAM},\ \bibinfo {year}
  {1981})\BibitemShut {NoStop}%
\bibitem [{\citenamefont {Camassa}\ and\ \citenamefont
  {Holm}(1993)}]{camassa1993integrable}%
  \BibitemOpen
  \bibfield  {author} {\bibinfo {author} {\bibfnamefont {R.}~\bibnamefont
  {Camassa}}\ and\ \bibinfo {author} {\bibfnamefont {D.~D.}\ \bibnamefont
  {Holm}},\ }\href@noop {} {\bibfield  {journal} {\bibinfo  {journal} {Phys.
  Rev. Lett.}\ }\textbf {\bibinfo {volume} {71}},\ \bibinfo {pages} {1661}
  (\bibinfo {year} {1993})}\BibitemShut {NoStop}%
\bibitem [{\citenamefont {Barthelemy}\ \emph {et~al.}(1985)\citenamefont
  {Barthelemy}, \citenamefont {Maneuf},\ and\ \citenamefont
  {Froehly}}]{barthelemy1985propagation}%
  \BibitemOpen
  \bibfield  {author} {\bibinfo {author} {\bibfnamefont {A.}~\bibnamefont
  {Barthelemy}}, \bibinfo {author} {\bibfnamefont {S.}~\bibnamefont {Maneuf}},
  \ and\ \bibinfo {author} {\bibfnamefont {C.}~\bibnamefont {Froehly}},\
  }\href@noop {} {\bibfield  {journal} {\bibinfo  {journal} {Opt. Commun.}\
  }\textbf {\bibinfo {volume} {55}},\ \bibinfo {pages} {201} (\bibinfo {year}
  {1985})}\BibitemShut {NoStop}%
\bibitem [{\citenamefont {Christodoulides}\ and\ \citenamefont
  {Joseph}(1988)}]{christodoulides1988discrete}%
  \BibitemOpen
  \bibfield  {author} {\bibinfo {author} {\bibfnamefont {D.~N.}\ \bibnamefont
  {Christodoulides}}\ and\ \bibinfo {author} {\bibfnamefont {R.~I.}\
  \bibnamefont {Joseph}},\ }\href@noop {} {\bibfield  {journal} {\bibinfo
  {journal} {Opt. Lett.}\ }\textbf {\bibinfo {volume} {13}},\ \bibinfo {pages}
  {794} (\bibinfo {year} {1988})}\BibitemShut {NoStop}%
\bibitem [{\citenamefont {Segev}\ \emph {et~al.}(1992)\citenamefont {Segev},
  \citenamefont {Crosignani}, \citenamefont {Yariv},\ and\ \citenamefont
  {Fischer}}]{segev1992spatial}%
  \BibitemOpen
  \bibfield  {author} {\bibinfo {author} {\bibfnamefont {M.}~\bibnamefont
  {Segev}}, \bibinfo {author} {\bibfnamefont {B.}~\bibnamefont {Crosignani}},
  \bibinfo {author} {\bibfnamefont {A.}~\bibnamefont {Yariv}}, \ and\ \bibinfo
  {author} {\bibfnamefont {B.}~\bibnamefont {Fischer}},\ }\href@noop {}
  {\bibfield  {journal} {\bibinfo  {journal} {Phys. Rev. Lett.}\ }\textbf
  {\bibinfo {volume} {68}},\ \bibinfo {pages} {923} (\bibinfo {year}
  {1992})}\BibitemShut {NoStop}%
\bibitem [{\citenamefont {Eisenberg}\ \emph {et~al.}(1998)\citenamefont
  {Eisenberg}, \citenamefont {Silberberg}, \citenamefont {Morandotti},
  \citenamefont {Boyd},\ and\ \citenamefont
  {Aitchison}}]{eisenberg1998discrete}%
  \BibitemOpen
  \bibfield  {author} {\bibinfo {author} {\bibfnamefont {H.~S.}\ \bibnamefont
  {Eisenberg}}, \bibinfo {author} {\bibfnamefont {Y.}~\bibnamefont
  {Silberberg}}, \bibinfo {author} {\bibfnamefont {R.}~\bibnamefont
  {Morandotti}}, \bibinfo {author} {\bibfnamefont {A.~R.}\ \bibnamefont
  {Boyd}}, \ and\ \bibinfo {author} {\bibfnamefont {J.~S.}\ \bibnamefont
  {Aitchison}},\ }\href@noop {} {\bibfield  {journal} {\bibinfo  {journal}
  {Phys. Rev. Lett.}\ }\textbf {\bibinfo {volume} {81}},\ \bibinfo {pages}
  {3383} (\bibinfo {year} {1998})}\BibitemShut {NoStop}%
\bibitem [{\citenamefont {Fleischer}\ \emph {et~al.}(2003)\citenamefont
  {Fleischer}, \citenamefont {Segev}, \citenamefont {Efremidis},\ and\
  \citenamefont {Christodoulides}}]{fleischer2003observation}%
  \BibitemOpen
  \bibfield  {author} {\bibinfo {author} {\bibfnamefont {J.~W.}\ \bibnamefont
  {Fleischer}}, \bibinfo {author} {\bibfnamefont {M.}~\bibnamefont {Segev}},
  \bibinfo {author} {\bibfnamefont {N.~K.}\ \bibnamefont {Efremidis}}, \ and\
  \bibinfo {author} {\bibfnamefont {D.~N.}\ \bibnamefont {Christodoulides}},\
  }\href@noop {} {\bibfield  {journal} {\bibinfo  {journal} {Nature}\ }\textbf
  {\bibinfo {volume} {422}},\ \bibinfo {pages} {147} (\bibinfo {year}
  {2003})}\BibitemShut {NoStop}%
\bibitem [{\citenamefont {Denschlag}\ \emph {et~al.}(2000)\citenamefont
  {Denschlag}, \citenamefont {Simsarian}, \citenamefont {Feder}, \citenamefont
  {Clark}, \citenamefont {Collins}, \citenamefont {Cubizolles}, \citenamefont
  {Deng}, \citenamefont {Hagley}, \citenamefont {Helmerson}, \citenamefont
  {Reinhardt} \emph {et~al.}}]{denschlag2000generating}%
  \BibitemOpen
  \bibfield  {author} {\bibinfo {author} {\bibfnamefont {J.}~\bibnamefont
  {Denschlag}}, \bibinfo {author} {\bibfnamefont {J.~E.}\ \bibnamefont
  {Simsarian}}, \bibinfo {author} {\bibfnamefont {D.~L.}\ \bibnamefont
  {Feder}}, \bibinfo {author} {\bibfnamefont {C.~W.}\ \bibnamefont {Clark}},
  \bibinfo {author} {\bibfnamefont {L.~A.}\ \bibnamefont {Collins}}, \bibinfo
  {author} {\bibfnamefont {J.}~\bibnamefont {Cubizolles}}, \bibinfo {author}
  {\bibfnamefont {L.}~\bibnamefont {Deng}}, \bibinfo {author} {\bibfnamefont
  {E.~W.}\ \bibnamefont {Hagley}}, \bibinfo {author} {\bibfnamefont
  {K.}~\bibnamefont {Helmerson}}, \bibinfo {author} {\bibfnamefont {W.~P.}\
  \bibnamefont {Reinhardt}},  \emph {et~al.},\ }\href@noop {} {\bibfield
  {journal} {\bibinfo  {journal} {Science}\ }\textbf {\bibinfo {volume}
  {287}},\ \bibinfo {pages} {97} (\bibinfo {year} {2000})}\BibitemShut
  {NoStop}%
\bibitem [{\citenamefont {Burger}\ \emph {et~al.}(1999)\citenamefont {Burger},
  \citenamefont {Bongs}, \citenamefont {Dettmer}, \citenamefont {Ertmer},
  \citenamefont {Sengstock}, \citenamefont {Sanpera}, \citenamefont
  {Shlyapnikov},\ and\ \citenamefont {Lewenstein}}]{burger1999dark}%
  \BibitemOpen
  \bibfield  {author} {\bibinfo {author} {\bibfnamefont {S.}~\bibnamefont
  {Burger}}, \bibinfo {author} {\bibfnamefont {K.}~\bibnamefont {Bongs}},
  \bibinfo {author} {\bibfnamefont {S.}~\bibnamefont {Dettmer}}, \bibinfo
  {author} {\bibfnamefont {W.}~\bibnamefont {Ertmer}}, \bibinfo {author}
  {\bibfnamefont {K.}~\bibnamefont {Sengstock}}, \bibinfo {author}
  {\bibfnamefont {A.}~\bibnamefont {Sanpera}}, \bibinfo {author} {\bibfnamefont
  {G.~V.}\ \bibnamefont {Shlyapnikov}}, \ and\ \bibinfo {author} {\bibfnamefont
  {M.}~\bibnamefont {Lewenstein}},\ }\href@noop {} {\bibfield  {journal}
  {\bibinfo  {journal} {Phys. Rev. Lett.}\ }\textbf {\bibinfo {volume} {83}},\
  \bibinfo {pages} {5198} (\bibinfo {year} {1999})}\BibitemShut {NoStop}%
\bibitem [{\citenamefont {Yang}\ \emph {et~al.}(1999)\citenamefont {Yang},
  \citenamefont {Malomed},\ and\ \citenamefont {Kaup}}]{yang1999embedded}%
  \BibitemOpen
  \bibfield  {author} {\bibinfo {author} {\bibfnamefont {J.}~\bibnamefont
  {Yang}}, \bibinfo {author} {\bibfnamefont {B.~A.}\ \bibnamefont {Malomed}}, \
  and\ \bibinfo {author} {\bibfnamefont {D.~J.}\ \bibnamefont {Kaup}},\
  }\href@noop {} {\bibfield  {journal} {\bibinfo  {journal} {Physical review
  letters}\ }\textbf {\bibinfo {volume} {83}},\ \bibinfo {pages} {1958}
  (\bibinfo {year} {1999})}\BibitemShut {NoStop}%
\bibitem [{\citenamefont {Yang}\ \emph {et~al.}(2001)\citenamefont {Yang},
  \citenamefont {Malomed}, \citenamefont {Kaup},\ and\ \citenamefont
  {Champneys}}]{yang2001embedded}%
  \BibitemOpen
  \bibfield  {author} {\bibinfo {author} {\bibfnamefont {J.}~\bibnamefont
  {Yang}}, \bibinfo {author} {\bibfnamefont {B.~A.}\ \bibnamefont {Malomed}},
  \bibinfo {author} {\bibfnamefont {D.~J.}\ \bibnamefont {Kaup}}, \ and\
  \bibinfo {author} {\bibfnamefont {A.~R.}\ \bibnamefont {Champneys}},\
  }\href@noop {} {\bibfield  {journal} {\bibinfo  {journal} {Mathematics and
  computers in Simulation}\ }\textbf {\bibinfo {volume} {56}},\ \bibinfo
  {pages} {585} (\bibinfo {year} {2001})}\BibitemShut {NoStop}%
\bibitem [{\citenamefont {Yagasaki}\ \emph {et~al.}(2005)\citenamefont
  {Yagasaki}, \citenamefont {Champneys},\ and\ \citenamefont
  {Malomed}}]{yagasaki2005discrete}%
  \BibitemOpen
  \bibfield  {author} {\bibinfo {author} {\bibfnamefont {K.}~\bibnamefont
  {Yagasaki}}, \bibinfo {author} {\bibfnamefont {A.~R.}\ \bibnamefont
  {Champneys}}, \ and\ \bibinfo {author} {\bibfnamefont {B.~A.}\ \bibnamefont
  {Malomed}},\ }\href@noop {} {\bibfield  {journal} {\bibinfo  {journal}
  {Nonlinearity}\ }\textbf {\bibinfo {volume} {18}},\ \bibinfo {pages} {2591}
  (\bibinfo {year} {2005})}\BibitemShut {NoStop}%
\bibitem [{\citenamefont {Oxtoby}\ and\ \citenamefont
  {Barashenkov}(2007)}]{oxtoby2007moving}%
  \BibitemOpen
  \bibfield  {author} {\bibinfo {author} {\bibfnamefont {O.~F.}\ \bibnamefont
  {Oxtoby}}\ and\ \bibinfo {author} {\bibfnamefont {I.~V.}\ \bibnamefont
  {Barashenkov}},\ }\href@noop {} {\bibfield  {journal} {\bibinfo  {journal}
  {Phys. Rev. E}\ }\textbf {\bibinfo {volume} {76}},\ \bibinfo {pages} {036603}
  (\bibinfo {year} {2007})}\BibitemShut {NoStop}%
\bibitem [{\citenamefont {Kevrekidis}(2009)}]{kevrekidis2009discrete}%
  \BibitemOpen
  \bibfield  {author} {\bibinfo {author} {\bibfnamefont {P.~G.}\ \bibnamefont
  {Kevrekidis}},\ }\href@noop {} {\emph {\bibinfo {title} {The discrete
  nonlinear Schr{\"o}dinger equation: mathematical analysis, numerical
  computations and physical perspectives}}},\ Vol.\ \bibinfo {volume} {232}\
  (\bibinfo  {publisher} {Springer Science and Business Media},\ \bibinfo
  {year} {2009})\BibitemShut {NoStop}%
\bibitem [{\citenamefont {Davis}\ \emph {et~al.}(1996)\citenamefont {Davis},
  \citenamefont {Miura}, \citenamefont {Sugimoto},\ and\ \citenamefont
  {Hirao}}]{davis1996writing}%
  \BibitemOpen
  \bibfield  {author} {\bibinfo {author} {\bibfnamefont {K.~M.}\ \bibnamefont
  {Davis}}, \bibinfo {author} {\bibfnamefont {K.}~\bibnamefont {Miura}},
  \bibinfo {author} {\bibfnamefont {N.}~\bibnamefont {Sugimoto}}, \ and\
  \bibinfo {author} {\bibfnamefont {K.}~\bibnamefont {Hirao}},\ }\href@noop {}
  {\bibfield  {journal} {\bibinfo  {journal} {Opt. Lett.}\ }\textbf {\bibinfo
  {volume} {21}},\ \bibinfo {pages} {1729} (\bibinfo {year}
  {1996})}\BibitemShut {NoStop}%
\bibitem [{\citenamefont {Gattass}\ and\ \citenamefont
  {Mazur}(2008)}]{gattass2008femtosecond}%
  \BibitemOpen
  \bibfield  {author} {\bibinfo {author} {\bibfnamefont {R.~R.}\ \bibnamefont
  {Gattass}}\ and\ \bibinfo {author} {\bibfnamefont {E.}~\bibnamefont
  {Mazur}},\ }\href@noop {} {\bibfield  {journal} {\bibinfo  {journal} {Nat.
  Photonics}\ }\textbf {\bibinfo {volume} {2}},\ \bibinfo {pages} {219}
  (\bibinfo {year} {2008})}\BibitemShut {NoStop}%
\bibitem [{\citenamefont {Szameit}\ and\ \citenamefont
  {Nolte}(2010)}]{szameit2010discrete}%
  \BibitemOpen
  \bibfield  {author} {\bibinfo {author} {\bibfnamefont {A.}~\bibnamefont
  {Szameit}}\ and\ \bibinfo {author} {\bibfnamefont {S.}~\bibnamefont
  {Nolte}},\ }\href@noop {} {\bibfield  {journal} {\bibinfo  {journal} {J.
  Phys. B: At. Mol. Opt. Phys.}\ }\textbf {\bibinfo {volume} {43}},\ \bibinfo
  {pages} {163001} (\bibinfo {year} {2010})}\BibitemShut {NoStop}%
\bibitem [{\citenamefont {Jensen}(1982)}]{jensen1982nonlinear}%
  \BibitemOpen
  \bibfield  {author} {\bibinfo {author} {\bibfnamefont {S.}~\bibnamefont
  {Jensen}},\ }\href@noop {} {\bibfield  {journal} {\bibinfo  {journal} {IEEE
  J. Quantum Electron.}\ }\textbf {\bibinfo {volume} {18}},\ \bibinfo {pages}
  {1580} (\bibinfo {year} {1982})}\BibitemShut {NoStop}%
\bibitem [{Edg()}]{EdgeSolAnimations}%
  \BibitemOpen
  \href {https://sites.psu.edu/leptos/2020/10/21/nonlinear-edge/} {}\bibinfo
  {note} {See Supplemental Movies~1-4 at\\
  {\color{blue}https://sites.psu.edu/leptos/2020/10/21/nonlinear-edge/}}\BibitemShut
  {NoStop}%
\end{thebibliography}%
\clearpage

\onecolumngrid
\appendix

\vspace{1.0cm}

\newcommand{\beginsupplement}{%
        \setcounter{equation}{0}
        \renewcommand{\theequation}{S\arabic{equation}}%
        \setcounter{figure}{0}
        \renewcommand{\thefigure}{S\arabic{figure}}%
     }
 
 \beginsupplement

\section{\large {Supplementary Information} \vspace*{0.3cm}}

\twocolumngrid

In the following sections of the Supplementary Materials, 
we discuss how the lifetime of the soliton-like edge states can be controlled by tuning the driving parameter $\delta$ and present further experimental details.

\begin{figure}[b!]
\center
\includegraphics[width=\linewidth]{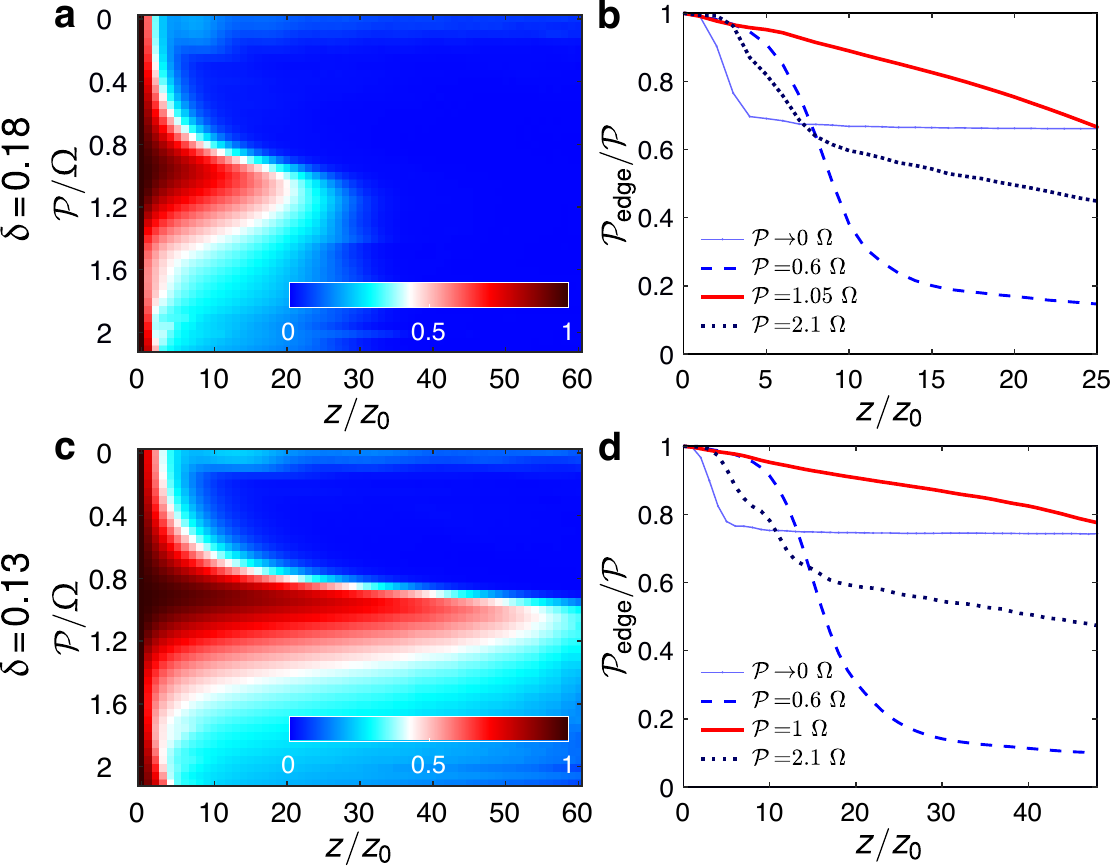}
\caption{{\bf IPR and power on the edge for two different values of $\delta$. (a, c)} The  variation of IPR, calculated after each driving period, as a function of renormalized power and propagation distance  for~$\delta\!=\!0.18$ and $0.13$, respectively. {\bf (b, d)} The variation of the fraction of total normalized power on the zig-zag edge, for different values of renormalized powers indicated on the figures, corresponding to $\delta\!=\!0.18$ and $0.13$, respectively.
}
\label{IPR_Overlap}
\end{figure}

\subsection{Finite lifetime of soliton-like edge states}
Because of the intrinsic gaplessness of the topological edge, confined traveling nonlinear waves will not in general live forever, and cannot therefore be called solitons (with the exception of the special case of embedded solitons).  In this section, we discuss the dependence of the lifetime of the soliton-like edge wavepackets on the dimerization parameter, $\delta$. In the main text, we presented numerical results showing unidirectional traveling soliton-like edge states, in Fig.~\ref{fig2}, for $\delta\!=\!0.18$. 
Supplementary Fig.~\ref{IPR_Overlap}~(a) shows the same variation of IPR, presented in Fig.~\ref{fig2}~(d), for a longer propagation distance. Up to approximately 20 driving cycles, the IPR exhibits a clear peak as a function of renormalized power at a given propagation distance. 
In this case, the fraction of light localized to the zig-zag edge $\mathcal{P}_{\text{edge}}/\mathcal{P}$ is plotted in Supplementary Fig.~\ref{IPR_Overlap}~(b) for four different renormalized powers $\mathcal{P}$, indicated in the figure. 
In the linear regime, $\mathcal{P}\!\rightarrow\!0$, the single-site input state largely overlaps $(\approx\!68\%)$ with the edge states, hence, some light penetrates into the bulk in the beginning ($0\!\le\!z/z_0\!\lesssim\!5$), and then, $\mathcal{P}_{\text{edge}}/\mathcal{P}$ remains unaltered, see the solid blue line in Supplementary Fig.~\ref{IPR_Overlap}~(b). At small power, $\mathcal{P}\!=\!0.6\Omega$, the wavepacket is able to nonlinearly couple to bulk modes and thus radiate away from the edge, see the blue dashed line in Supplementary Fig.~\ref{IPR_Overlap}~(b).  Furthermore, this is below the power threshold for the formation of the soliton-like wavepacket, meaning that there is still significant diffraction along the edge. At a specific value of the renormalized power, $\mathcal{P}\!=\!1.05\Omega$, the radiation into the bulk reduces (see the solid red line), and importantly the state propagates long distance without significant diffraction along the edge, corresponding to the long-lived soliton-like object, and hence a peak in the IPR. At higher power, $\mathcal{P}\!=\!2.1\Omega$, the radiation increases again (see the dotted dark blue line) and the state diffracts along the edge during propagation.

The radiation rate of the soliton-like edge state increases with increasing $\delta$. Figure ~\ref{IPR_Overlap}~(c, d) presents the same results as Fig.~\ref{IPR_Overlap}~(a, b) except with a lower value of $\delta\!=\!0.13$.
In this case, the clear peak in the IPR can be observed up to a longer propagation distance $z\!\approx 55z_0$. The result in Fig.~\ref{IPR_Overlap}~(d) is qualitatively similar to Fig.~\ref{IPR_Overlap}~(b), however, the important difference is that the dynamics is slower at this smaller value of $\delta$ -- the soliton-like edge state (the solid red line) has a comparatively longer lifetime. 
In the limit of $\delta\!\rightarrow\!0$, the edge spectrum has linear dispersion and the bulk is dispersionless. In this case, nondiffracting unidirectional edge transport is observed for arbitrarily long propagation distance in the limit of zero nonlinearity, $\mathcal{P}\!\rightarrow\!0$.

\begin{figure*}[t!]
\center
\includegraphics[width=13.5cm]{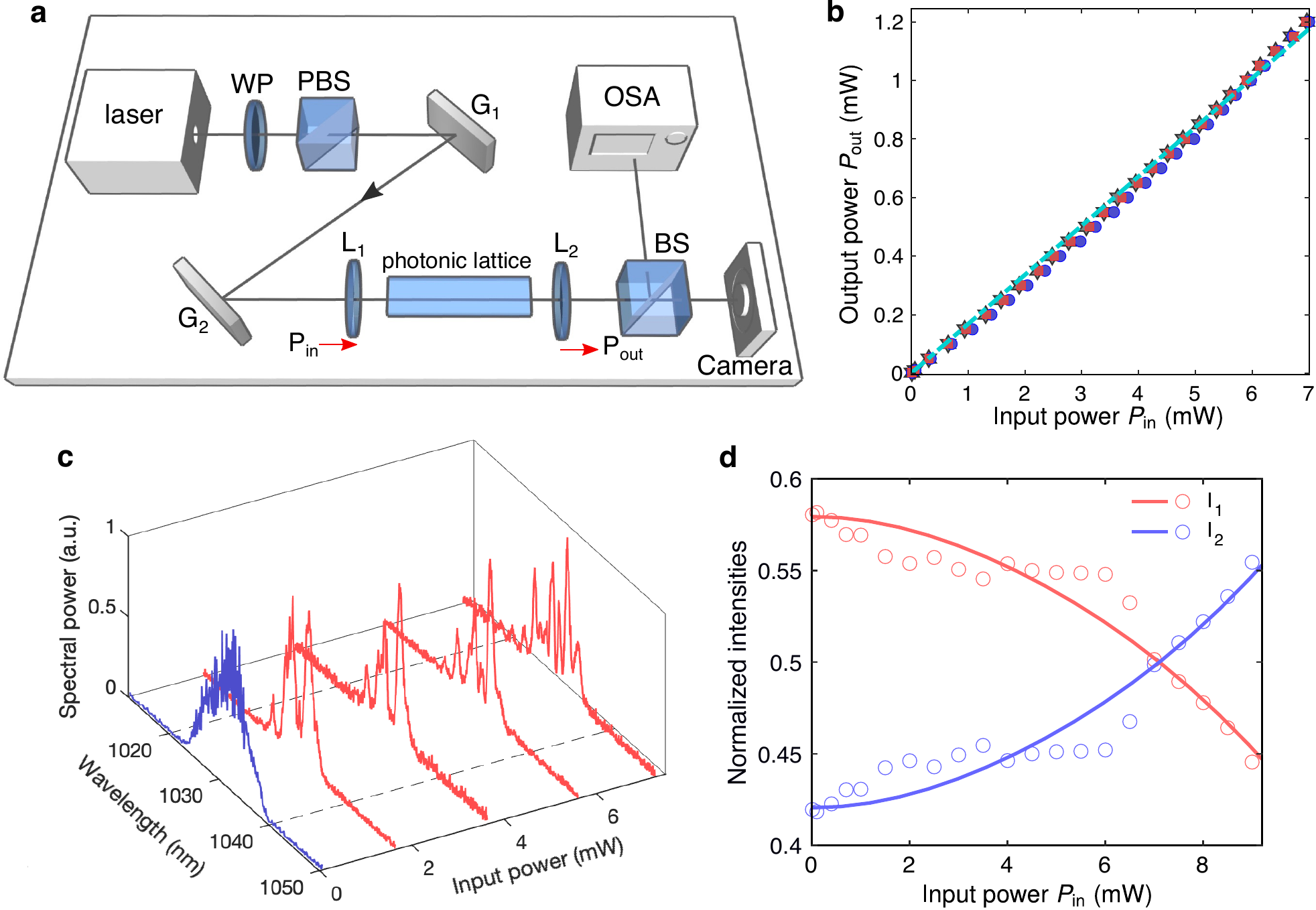}
\caption{{\bf Nonlinear characterization. (a)} Simplified schematic of the nonlinear characterization setup. Here WP is a half-waveplate, PBS is a polarizing beam splitter, G$_{1,2}$ is a parallel grating pair, L$_{1,2}$ are air-coated convex lenses, and BS is a beam splitter. Laser pulse trains at $1030\! \pm 4$ nm wavelength, $5$ kHz pulse repetition rate and $260$ fs pulse duration were generated using a
Yb-doped fiber laser (Menlo BlueCut) system. The parallel grating pair is used to temporally stretch (and down-chirp) the pulses to $2$ ps. The intensity patterns at the output of the photonic lattices are imaged on a CCD camera and the optical spectrum (intensity vs wavelength) is measured using an optical spectrum analyzer (OSA). {\bf (b)} The variation of average output power $P_{\text{out}}$ with the input power $P_{\text{in}}$. The linear variation implies that nonlinear losses due to multi-photon absorption processes can be ignored in our experiments. Here three data sets are shown, and the dashed line indicates a linear fit. {\bf (c)} Self-phase modulation (SPM) induced by Kerr nonlinearity. Here the normalized spectral powers are shown for five sets of average input powers. The spectral width in the linear regime (shown in blue) is $\approx \! 8$ nm (FWHM). At higher powers, the spectrum changes due to SPM, however, the maximal spectral width after 76 mm of propagation was measured to be less than $20$ nm for the maximal required nonlinearity. {\bf (d)} The renormalized power $\mathcal{P}$ is estimated by characterizing a nonlinear directional coupler. The data in red (blue) are the measured normalized intensity $I_1$ at the output of waveguide $1$ $(2)$ (light was launched into waveguide $1$).  The solid lines were obtained by solving Eq.~\ref{nl_coupler} and fitting
$\mathcal{P}$ at the input to be $0.046$ mm$^{-1}$ for unit input power in mW.
}
\label{setup}
\end{figure*}

\subsection{More experimental details}  

{\it Fabrication}: Topological photonic lattices consisting of periodically-modulated 
single-mode optical waveguides were created using femtosecond (fs) laser writing, an on-chip device fabrication technique~\cite{davis1996writing, gattass2008femtosecond, szameit2010discrete}. 
This technique allows us to precisely control the waveguide paths in a three-dimensional geometry, which is crucial for realizing the nontrivial topology considered in this work. 
Laser pulses at $1030\! \pm4$ nm wavelength, $500$ kHz pulse repetition rate, and $260$ fs pulse width were generated using a commercially available Yb-doped (Menlo BluCut) fiber laser system. Each waveguide in the lattice was inscribed by translating a borosilicate (Corning Eagle XG) glass substrate -- which was mounted on a high-precision $x$-$y$-$z$ (Aerotech) translation stages -- once through the focus of the femtosecond laser beam. The laser pulse energy and translation speed of fabrication were optimized to obtain low-loss single mode optical waveguides near the $1030$nm wavelength. The maximal insertion (propagation + bend + input coupling) loss for the entire lattice was measured to be $7.5$ dB, and no significant polarization-dependent loss was detected in our experiments.

{\it Nonlinear characterization:} Fig.~\ref{setup}~(a) presents a simplified schematic of the nonlinear characterization setup. We used laser pulses at $5$ kHz repetition rate generated by the Menlo BlueCut system. The average input power can be tuned in our experiments by using a combination of a half-waveplate and a polarizing beam splitter. A parallel grating pair is used to temporally stretch (and down-chirp) the pulses to $2$~ps.
To gauge the nonlinear loss due to multi-photon absorption, we measured the average output power as a function of average input power for all nonlinear characterizations, and a linear variation was observed, as shown in Supplementary Fig.~\ref{setup}~(b). This linear variation of $P_{\text{out}}$ with the input power $P_{\text{in}}$ implies that nonlinear losses can be ignored in our experiments.

As mentioned in the main text, the temporal shape of the laser pulses can cause undesired effects such as self-phase modulation (i.e., generation of new wavelengths) and chromatic dispersion. To access self-focusing nonlinearity with a minimal self-phase modulation, we use temporally stretched and down-chirped laser pulses. In this situation, a maximal spectral width of $<\!20$ nm was observed in our experiments, see Fig.~\ref{setup}~(c). In this wavelength range, the evanescent coupling $J$ only varies by $\Delta J/J\!=\!\pm 4$ \%, which is of the order of the unavoidable small disorder present in the lattice. Hence, we can safely ignore the effects of self-phase modulation in our experiments. The effects of chromatic dispersion is negligibly small for the maximal propagation distance ($76$ mm) considered in our experiments~\cite{mukherjee2020observation}.

{\it Estimation of $\mathcal{P}$}: The renormalized power $\mathcal{P}$ was experimentally calibrated by characterizing a two-waveguide directional coupler with a known coupling strength and linear loss coefficient. The paraxial propagation of light in a nonlinear directional coupler is governed by~\cite{jensen1982nonlinear}
\begin{eqnarray}
\label{nl_coupler}
i \partial_z \Phi_{1,2}\!=\!-J \Phi_{2, 1}- |\Phi_{1,2}|^2 \Phi_{1,2} -i\alpha  \Phi_{1,2},
\end{eqnarray}
where $J$ is the coupling strength, $\alpha$ is a measure of linear loss, and $|\Phi_{1,2}|^2$ is proportional to the optical power at waveguide $1$ and $2$, respectively. It should be mentioned that $\mathcal{P}\!=\!(|\Phi_{1}|^2+|\Phi_{2}|^2)$  is not a conserved quantity during propagation when optical losses are present -- here, we experimentally estimate  $\mathcal{P}(z\!=\!0)$.
In the experiments, we launch light at waveguide $1$; the variation of normalized output intensities $|\Phi_{1,2}|^2/(|\Phi_{1}|^2+|\Phi_{2}|^2)$ with average input power is presented in Fig.~\ref{setup}~(d). The solid lines were obtained by solving Eq.~\ref{nl_coupler} and fitting
$\mathcal{P}$ at the input to be $0.046$ mm$^{-1}$ for unit input power in mW.

\subsection{Description of the supplementary movies~\cite{EdgeSolAnimations}}

Movie~1-3: Propagation of a single-site input wavepacket on the edge of the periodically-driven lattice $(\delta\!=\!0.18)$ with three different renormalized powers $\mathcal{P}$, indicated on each movie. Unlike Fig.~\ref{fig2}~(a-c) in the main text, the full two-dimensional lattice has been shown in these movies.
For $\mathcal{P}\!\rightarrow \!0$ (Movie~1) and $\mathcal{P}\!=\!2.1 \Omega$ (Movie~3)
the input state spreads out along the edge as well as penetrates the bulk. At a certain intermediate power value ($\mathcal{P}\!=\!1.05 \Omega$, Movie~2), the input state propagates unidirectionally while maintaining its shape up to a long propagation distance.\\

Movie~4:  Experimentally measured intensity patterns (left) and calculated inverse participation ratio (right) at propagation distance $z\!=\!2z_0$ for the driving parameter $\delta\!=\!0.26$ -- a clear peak in the IPR is visible as a function of input power. The input state propagates two unit cells along the bottom edge with minimal spreading when the average input power is near $P_{\text{in}}\!=\!5.1$ mW. The white circle indicates the lattice site where light was launched at the input. Each image is normalized, and the field of view is smaller than the lattice size.

\clearpage
\end{document}